\input harvmac
\input epsf

%
\def\figin{\epsfcheck\figin}\def\figins{\epsfcheck\figins}
\def\epsfcheck{\ifx\epsfbox\UnDeFiNeD
\message{(NO epsf.tex, FIGURES WILL BE IGNORED)}
\gdef\figin##1{\vskip2in}\gdef\figins##1{\hskip.5in}
\else\message{(FIGURES WILL BE INCLUDED)}%
\gdef\figin##1{##1}\gdef\figins##1{##1}\fi}
\def\DefWarn#1{}
\def\figinsert{\goodbreak\midinsert}
\def\ifigc#1#2#3{\DefWarn#1\xdef#1{\the\figno}
\writedef{#1\leftbracket \the\figno}
\figinsert\figin{\centerline{#3}}\medskip
\centerline{\vbox{\baselineskip\footskip
\centerline{\footnotefont{\bf Fig.~\the\figno:} #2}}}
\bigskip\endinsert\global\advance\figno by1}
\def\ifig#1#2#3{\DefWarn#1\xdef#1{\the\figno}
\writedef{#1\leftbracket \the\figno}
\figinsert\figin{\centerline{#3}}\medskip
\centerline{\vbox{\baselineskip\footskip
\advance\hsize by -1truein\noindent
\footnotefont{\bf Fig.~\the\figno:} #2}}
\bigskip\endinsert\global\advance\figno by1}
%
\def\Title#1#2{\rightline{#1}\ifx\answ\bigans\nopagenumbers\pageno0
\vskip0.5in
\else\pageno1\vskip.5in\fi \centerline{\titlefont #2}\vskip .3in}

\font\caps=cmcsc10

\noblackbox
\parskip=1.5mm

  
\def\npb#1#2#3{{\it Nucl. Phys.} {\bf B#1} (#2) #3 }
\def\plb#1#2#3{{\it Phys. Lett.} {\bf B#1} (#2) #3 }
\def\prd#1#2#3{{\it Phys. Rev. } {\bf D#1} (#2) #3 }
\def\prl#1#2#3{{\it Phys. Rev. Lett.} {\bf #1} (#2) #3 }
\def\mpla#1#2#3{{\it Mod. Phys. Lett.} {\bf A#1} (#2) #3 }

\def\cmp#1#2#3{{\it Commun. Math. Phys.} {\bf #1} (#2) #3 }

\def\bbg#1{{\tt gr-qc/#1}}
\def\bb#1{{\tt hep-th/#1}}

\def\jhep#1#2#3{{\it J. High Energy Phys.} {\bf #1} (#2) #3 }


           \def\CO{{\cal O}}

\def\CN{{\cal N}}


\def\dj{\hbox{d\kern-0.347em \vrule width 0.3em height 1.252ex depth
-1.21ex \kern 0.051em}}

\def\half{{1\over 2}\,}

\def\Tr{{\rm Tr\,}}

\def\ket{\rangle}
\def\bra{\langle}

\def\pt{\partial}

\def\Vol{{\rm Vol}({\bf S}^{8-p}) }  
\def\ls{\ell_s}
\def\Xbh{X_{\rm bh}}
\def\Xvac{X_{\rm vac}}
\def\Xhag{X_{\rm Hag}}
         

\lref\rik{S.S. Gubser, I.R. Klebanov and A.W. Peet, \prd{54}{1996}{3915\semi}
I.R. Klebanov, \npb{496}{1997}{217\semi}
S.S. Gubser, I.R. Klebanov and A.A. Tseytlin, \npb{499}{1997}{217\semi} 
S.S. Gubser and I.R. Klebanov, \plb{413}{1997}{41.}}
\lref\rpol{A.M. Polyakov, {\it ``String Theory And Quark Confinement",} 
\bb{9711002.}}
\lref\rsup{G.W.  Gibbons and P.K. Townsend, \prl{71}{1993}{5223\semi}
G.W. Gibbons, G.T. Horowitz and P.K. Townsend, {\it Class. Quant. Grav.}
{\bf 12} (1995) 297.} 
 \lref\rhoo{G. T. Horowitz and H. Ooguri, \prl{80}{1998}{4116,}   
 \bb{9802116.}}
\lref\rgo{D.J. Gross and H. Ooguri, {\it ``Aspects of Large $N$ Gauge
Dynamics as Seen by String Theory,"} \bb{9805129.}}  
\lref\rholog{G. 't Hooft, {\it ``Dimensional Reduction In Quantum Gravity"},
in Salamfest, Trieste, 1993, p. 284, \bbg{9310026\semi}
J.D. Bekenstein, \prd{49}{1994}{1912\semi}
L. Susskind,  J. Math. Phys. {\bf 36} (1995)
6377, \bb{9409089.} }
\lref\rli{M. Li, {\it ``Evidence for Large N Phase Transition in $N=4$
Super Yang--Mills Theory at Finite Temperature"}, \bb{9807196.}} 
\lref\radm{F. Tangherlini, {\it Nuovo Cimento} {\bf 77} (1963) 636\semi
R. Myers and M. Perry, {\it Ann. Phys.} {\bf 172} (1986) 304.}    
\lref\rws{L. Susskind and E. Witten, {\it ``The Holographic Bound
In Anti-de Sitter Space,"} \bb{9805114.}} 
\lref\rwituno{E. Witten, {\it ``Anti De Sitter Space and Holography"},
\bb{9802150.} }
\lref\rwitdos{E. Witten, {\it ``Anti-de Sitter Space, Thermal Phase Transition,
And Confinement In Gauge Theories"}, \bb{9803131.}}
\lref\rmalda{ J. Maldacena, {\it ``The Large $N$ Limit Of Superconformal Field
Theories And Supergravity"}, \bb{9711200.}}
\lref\rtelaviv{N. Itzhaki,
 J. Maldacena, J. Sonnenschein, and S. Yankielowicz,  
\prd {58}{1998}{046004,} \bb{9802042.}}
\lref\rgkp{S.S. Gubser, I.R. Klebanov and A.M. Polyakov,  
\plb{428}{1998}{105,} 
\bb{9802109.}}    
\lref\rgkse{A. Giveon, D. Kutasov and N. Seiberg, {\it ``Comments on
String Theory on $AdS_3$",} \bb{9806194.}} 
\lref\rloopt{  
 A. Brandhuber, N. Itzhaki, J. Sonnenschein and S. Yankielowicz,  
{\it``Wilson Loops in the Large $N$ Limit at Finite Temperature,"}
 \bb{9803137\semi} 
 S-J. Rey,   S. Theisen and J-T. Yee, 
\npb{527}{1998}{171,}    \bb{9803135.}}   
\lref\ratwitt{J. Atick and E. Witten,  
\npb{310}{1988}{291.}}  
\lref\rgkp{S.S. Gubser, I.R. Klebanov and A. M. Polyakov, {\it ``Gauge Theory
Correlators From Noncritical String Theory"}, \bb{9802109.}}
\lref\rcorrp{G.T. Horowitz and J. Polchinski, \prd{55}{1997}{6189,} 
\bb{9612146.}}  
\lref\rfrac{S. Das and S. Mathur, \plb{365}{1996}{79,} \bb{9601152\semi}  
J.M. Maldacena and L. Susskind, \npb{475}{1996}{679,}  
\bb{9604042.}}  
\lref\rgl{R. Gregory and R. Laflamme, \prl{70}{1993}{2837,} \bb{9301052.}}
\lref\rkt{I.R. Klebanov and A.A. Tseytlin, \npb{475}{1996}{164,} 
\bb{9604089.}}  
\lref\rmaldafive{J.M. Maldacena,   \npb{477}{1996}{168,}  
\bb{9605016.}}  
\lref\rgreenbs{M.B. Green, \plb{266}{1991}{325,}  \plb{329}{1994}{435,} 
\bb{9403040.}}  
\lref\rholog{G. 't Hooft, {\it ``Dimensional Reduction In Quantum Gravity"},
in Salamfest, Trieste, 1993, p. 284, \bbg{9310026\semi}
J.D. Bekenstein, \prd{49}{1994}{1912\semi}
L. Susskind,  J. Math. Phys. {\bf 36} (1995)
6377, \bb{9409089.} }
\lref\rigor{S. S. Gubser, I. R. Klebanov and A. W. Peet,  \prd{54}{1996}{3915,}
\bb{9602135.\semi}
I. R. Klebanov, \npb {496}{1997}{231,} \bb{9702076\semi}
S. S. Gubser, I. R. Klebanov and A.A. Tseytlin, \npb {499}{1997}{217,} \bb{9793940
\semi}
S. S. Gubser and I. R. Klebanov, \plb {413}{1997}{41,} \bb{9708005.}}
\lref\rgkts{S.S. Gubser, I.R. Klebanov and A.A. Tseytlin, {\it `Coupling
Constant Dependence in the Thermodynamics of $N=4$ Supersymmetric Yang--Mills
Theory",} \bb{9805156.}} 
\lref\ram{S.S. Gubser, I.R. Klebanov and A.W. Peet, \prd{54}{1996}{3915,} 
\bb{9602135.}} 
\lref\rsmal{J. Maldacena and A. Strominger, \jhep {12}{1997}{008,}  
 \bb{9710014.}}
\lref\rhawkp{S. Hawking and D. Page, \cmp {78B}{1983}{577.}}
\lref\rhawkg{ G. W. Gibbons and S. Hawking, \prd{15}{1977}{2752.}}
\lref\rbhprobes{ M. R. Douglas, J. Polchinksi and A. Strominger,      
  \jhep {12}
{1997} {003,}
 \bb{9703031.}}
\lref\rkutsei{D. Kutasov and N. Seiberg, \npb{358}{1991}{600.}}
\lref\rceff{D. Kutasov, \mpla{7}{1992}{2943\semi}
 E. Hsu and D. Kutasov, \npb{396}{1993}{693.}}
\lref\rthooftb{G. 't Hooft, \npb{256}{1985}{727.}}
\lref\rsuglum{L. Susskind and J. Uglum, \prd{50}{1994}{2700.}}
\lref\rgreenbk{T. Banks and M.B. Green, 
\jhep{05}{1998}{002,}  
\bb{9804170\semi}
M. Bianchi, M.B. Green, S. Kovacs and G. Rossi,   
{\it ``Instantons in Supersymmetric Yang--Mills and D-Instantons in
IIB Superstring Theory",} \bb{9807033.}}  
\lref\rwitot{E. Witten, {\it ``Theta Dependence In The Large $N$ Limit
Of Four Dimensional Yang--Mills Theories",} \bb{9807109.}}
\lref\rsuw{L. Susskind and E. Witten, {\it ``The Holographic Bound in
Anti-de Sitter Space"}, \bb{9805114.}}
\lref\rbrs{J.L.F. Barb\'on, \plb {339}{1994}{41}, \bb{9406209.}}
\lref\rma{M.A. V\'azquez-Mozo, \plb{388}{1996}{494,} \bb{9607052.}} 
\lref\rloop{J.M. Maldacena, \prl{80}{1988}{4859,} \bb{9803002\semi} 
S-J Rey and J. Yee, {\it ``Macroscopic Strings as Heavy Quarks in Large N
Gauge Theory and Anti-de Sitter Supergravity",}  \bb{9803001.}}  
\lref\rsen{
L. Susskind, {\it ``Some Speculations about Black Hole Entropy in String Theory",}
 \bb{9309145\semi}   
A. Sen, \mpla{10}{1995}{2081,} \bb{9504197\semi} 
G. Horowitz, {\it ``The Origin of Black Hole Entropy in String Theory",} 
\bbg{9604051.}} 
\lref\rigg{S.S. Gubser, A. Hashimoto, I.R. Klebanov and M. Krasnitz, 
\npb{526}{1998}{393,} \bb{9803023.}}  
\lref\rdalw{S. de Alwis, {\it ``Supergravity, the DBI Action and
Black Hole Physics"}, \bb{9804019.}} 
\lref\rdp{G. Horowitz and A. Strominger, \npb{360}{1991}{197\semi}
M.J. Duff, R.R. Khuri and J.X. Lu,  {\it Phys. Rep.} 
{\bf 259} (1995) 213, \bb{9412184.} } 
\lref\rbr{J.L.F. Barb\'on and E. Rabinovici, {\it ``Extensivity
Versus Holography in Anti-de Sitter spaces"}, \bb{9805143\semi}
E. Rabinovici, 
 {\tt http://www.itp.ucsb.edu/online/strings98/rabinovici/}}          
\lref\rfc{S. Frautschi, \prd{3}{1971}{2821\semi} 
R.D. Carlitz, \prd{5}{1972}{3231.}} 
\lref\rvort{I.I. Kogan, {\it JETP. Lett.} {\bf 45} (1987) 709\semi
B. Sathiapalan, \prd{35}{3277.}}     
\lref\rbkt{V. Berezinski, {\it JETP. Lett.} {\bf 32} (1971) 493\semi
J. Kosterlitz and D. Thouless, {\it J. Phys. } {\bf C6} (1973) 1181.}     
\lref\rmave{N. Mavromatos, {\it ``World-sheet Defects, Strings and
Quark Confinement,"} $\;$ 
 \bb{9803189.}} 
\lref\rtelavd{A. Brandhuber, N. Itzhaki, J. Sonnenschein and S. Yankielowicz,
 \jhep{06}{1998}{001}, \bb{9803263.}}  
\lref\ropvol{G. W. Gibbons and M. J. Perry, {\it Proc. R. Soc. Lond.} {\bf A358},
(1978), 467 \semi
J.S. Dowker and G. Kennedy, {\it J. Phys.} {\bf A11} (1978) 895.  }
\lref\rlupo{H. Lu, S. Mukherji, C. Pope and J. Rahmfeld, \plb{389}{1996}{248,}
 \bb{9604127.}} 
\lref\rbfks{T. Banks, W. Fishler, I.R. Klebanov and L. Susskind, 
\prl{80}{1998}{226,}   
\bb{9709091,} \jhep{01}{1998}{008,}   
\bb{9711005.}}     
\lref\rhmar{G. Horowitz and E. Martinec, \prd{57}{1998}{4935,}  
\bb{9710217.}} 
\lref\rgwi{D.J. Gross and E. Witten, \prd{21}{1980}{446.}}
\lref\rpolh{G. Polhemus, \prd{56}{1998}{2202,}  
\bb{9612130.}} 
\lref\rkinst{I.I. Kogan and G. Luz\'on, {\it `D-Instantons on the
Boundary",} \bb{9806197.}}    
\lref\rsgwi{L. Susskind, {\it `Matrix Theory Black Holes and the 
Gross--Witten Transition",} \bb{9805115.}} 
\lref\rss{A. Sen, {\it Adv. Theor. Math. Phys.} {\bf 2} (1998) 51,      
\bb{9709220\semi}     
N. Seiberg, \prl{79}{1997}{3577,}   
\bb{9710009.}}  
\lref\rchin{C.-S. Chu, P.M. Ho and Y.-Y. Wu,
 {\it ``D-Instanton in $AdS_5$ and Instanton in $SYM_4$",} \bb{9806103.}}
\lref\rbir{D. Birmingham, {\it ``Topological Black Holes in Anti- de Sitter
Space,"} \bb{9808032.}} 


\line{\hfill CERN-TH/98-206}
\line{\hfill OUTP-98-48P}
\line{\hfill {\tt hep-th/9809033}}
 
\Title{\vbox{\baselineskip 12pt\hbox{}
 }}
{\vbox {\centerline{On Stringy Thresholds in SYM/AdS Thermodynamics}  
}}
 
\vskip 0.1cm
 
\centerline{$\quad$ {{\caps J. L. F. Barb\'on}$^{\,\rm a,}$
\foot{{\tt barbon@mail.cern.ch},
$^2\,\rm ${\tt i.kogan@physics.ox.ac.uk},
$^3${\tt eliezer@vxcern.cern.ch  }},
{\caps I. I. Kogan}$^{\,\rm b,2}$ and
{\caps E. Rabinovici}$^{\,{\rm a},{\rm c},3}$
 }}
\vskip0.1cm 
 
\centerline{{\sl $^{\rm a}$Theory Division, CERN}}
\centerline{{\sl CH-1211, Geneva 23, Switzerland}}

\vskip0.1cm
\centerline{{\sl $^{\rm b}$Theoretical Physics, Department of Physics}}
\centerline{{\sl 1 Keble Road, Oxford, OX1 3NP, UK}}
 
\vskip0.1cm

\centerline{{\sl $^{\rm c}$Racah Institute of Physics}}
\centerline{{\sl The Hebrew University}}
\centerline{{\sl  Jerusalem 91904, Israel}}
\vskip0.1cm

We  consider
aspects of the role of stringy scales and Hagedorn temperatures in
the correspondence between various field theories and AdS-type spaces. 
 The boundary theory is set on
a toroidal world-volume to enable small scales
to appear in
the supergravity backgrounds also for low field-theory temperatures.
We find that thermodynamical considerations tend to  favour
 background manifolds with
no string-size
characteristic 
scales. The gravitational dynamics censors the reliable exposure of Hagedorn
physics on the supergravity side, and  the
 system does not allow the study
 of the Hagedorn scale by low-temperature field theories.
These results are obtained following some heuristic assumptions on
the character of stringy modifications to the gravitational
backgrounds.  
A rich
phenomenology appears on the supergravity side, 
with different string backgrounds 
 dominating in  different regions, which should have field-theoretic
consequences.  Six-dimensional 
world volumes turn out to be borderline cases from several points of view. For
lower dimensional world-volumes, a fully holographic behaviour is exhibited to
order $1/N^2$,
 and open strings in their presence are 
found to have a thermodynamic Hagedorn 
behaviour similar to that of closed strings in 
flat space.

\vskip 0.1cm
 
\noindent
 


\Date{09/98}


\newsec{Introduction}
The suggestion of a correspondence between a class of effective conformal
 field theories and  special, anti- de Sitter  (AdS), string backgrounds,
\refs\rmalda, has 
allowed the  study of some  basic properties of gauge theories in the strong
coupling limit\foot{Earlier
work in this direction includes \refs\rik, \refs\rsup, \refs\rbhprobes.}. One
 may
also attempt to turn the tables around and investigate the possibility that 
the correspondence sheds some light on outstanding problems in gravity and
string theory. We would like to investigate the extent to which physics at
small (i.e. string) distance scales emerging on the supergravity side can be
understood in terms of gauge dynamics.  One such issue is the nature of the
Hagedorn transition. We will report here on a search for an answer to that
question. Actually, something has already been learned about the nature of
gravity at the very first stage of setting up the correspondence. The
correspondence can work only if the theory of gravity is allowed to contain at
the same time configurations of different topologies. This is vindicated by
studying thermodynamical aspects of the correspondence.  The thermodynamics on
the Conformal Field Theory side, (CFT), 
 was studied in \refs\ram, and  \refs\rwituno, \refs\rwitdos,  
 on the basis of the large $N$  CFT/AdS conjectured correspondence.       
For strong coupling the CFT thermodynamics is obtained by evaluating the 
gravitational thermodynamics in an AdS background. When several masterfields
may contribute on the supergravity side, one is instructed to take all of them
into account, \refs\rwituno, 
 and  this is needed to fully acount for thermodynamical properties 
of the CFT.

 In
particular, when the spatial world-volume of the
 four-dimensional gauge
theory has the topology of ${\bf S}^3$, there are two possible bulk manifolds 
ending on the same world-volume boundary.
One is an AdS-like vacuum
which we shall  denote $\Xvac$,  while the other is a black-hole
like manifold denoted $X_{\rm bh}$. Both are taken into account in calculations
performed on the supergravity side. The transition between them at some finite
temperature expresses a first-order phase transition following from the
occurence of a sort of gravitational collapse, along the lines of
\refs\rhawkp.  Technically, it is based on the balance of the gravitational
free energy between the vacuum AdS space with a radiation thermal ensemble
defined on it, or an AdS black hole in equilibrium with the radiation at
the same temperature. Such a  transition is indeed of 
first-order in the large $N$ limit (the classical limit from the gravity
point of view), and it has been established for the case of spherical topology.
The feature of several masterfields contributing has been responsible
for the holographic \refs\rholog\
 nature of the mapping at the classical \refs\rwituno, 
\refs\rwitdos,  
and one-loop level, \refs\rbr. It was also the gravitational manifestation
of (as well as the motivation to search for) a large $N$ phase transition
at finite sphere volume in gauge theory. It will turn out that it also 
influences the extent to which one may probe the nature of the Hagedorn
transition with these methods.

 The above large $N$ phase transition is
not driven by the emergence of light states involved in the onset
of the Hagedorn phenomena.
In order to search for such states one needs to first address two issues.
The first concerns the identification of the possibility of the emergence of
light string states (states which could lead to a Hagedorn instability) by
analysing a given supergravity background. These we expect to occur whenever
such a manifold has a string-scale circumference in the Euclidean time direction.
  The 
second concerns the estimate of the gauge theory temperature,  
dual to a given supergravity configuration. The length of the euclidean time
circumference depends in general on the value of the spatial coordinates.
We estimate the dual gauge theory temperature by considering the circumference
at those points for which the radial AdS coordinate is of the scale
 appropriate to the
 AdS curvature (which is equal to the scale of the complementary
compact manifold).
From this point of view a  search \refs\rbr\
 for those states in the case of ${\bf S}^3$
 topology has indicated the existence of a `Hagedorn censorship'.
The manifold $\Xvac$ is found to contain non-contractible
thermal loops whose length is of the
order of the string scale, $\ell_s \sim \sqrt{\alpha'}$,  
 only when the temperature of the gauge theory
itself is of order
\eqn\tha{T_{\rm gauge} = R^{-1} (g_s N)^{1/4} = 1/\ls,}
 where
$g_s$ is proportional to the string coupling, and $R$, the radius
of the spatial ${\bf S}^3$, was set at $\ls(g_s N)^{1/4}$ to estimate the gauge
theory temperature.

 However,
 at temperatures of order $T_c \sim R^{-1}$, much smaller than  \tha\
in the supergravity regime $g_s N \gg 1$,  
 the manifold $\Xbh$ takes over the thermodynamical description
and its horizon shields any string-length non-contractible thermal
 loops. This inability of
the gauge theory to allow a peek into the Hagedorn regime we have termed
Hagedorn censorship. In principle, string theory should serve as our guide to
what the exact correspondence
is and to
its possible limitations, \refs\rgkse.
 As long as this is not resolved in general, we
explore 
several possibilities. The first is that  the correspondence is valid at string
scale temperatures. In that case  one could  be learning  something about the
string theory, i.e. that there is no Hagedorn transition in
this string background, perhaps as a result of a decay and collapse of the
towers of excited states. Another possibility, which we check in this paper, 
is that the Hagedorn transition may be exposed if the world volume of the
boundary gauge theory is modified so that the supergravity background
geometries contain string scales also for low gauge theory temperatures. We
also probe the possibility that at string scales the corresponding systems  
change so as to allow for the expression of string scales on both sides. A
rather rich phenomenology will unfold and at the same time the  theory will
continue to resist devulging reliable information on a Hagedorn transition.

 In section 2  we choose the topology of the world volume so that small scales
are present also for low gauge theory temperatures. This is done by  setting 
up the gauge theory in a different spatial topology: a three-torus   
${\bf T}^3$ . Manifolds corresponding to the $ \Xvac$ and $\Xbh$ ones defined
for 
the ${\bf S}^3$
topology exist also for this case  (see \refs\rbir, and references
therein,  for more complicated
topolgies in the case of two-dimensional horizons).  The manifolds are similar in 
some ways but not identical, in particular
$\Xbh$ is well defined for this topology at all temperatures and 
non-contractible 
thermal (as well as spatial) loops of string-scale length are present
in $\Xvac$ for arbitrarily low gauge-theory temperatures. This would thus seem
an appropriate arena in which to attempt to circumvent the Hagedorn censorship.
The discussion  can be carried out    also for
 more general ${\bf T}^p$ spatial manifolds,
 obtained from the
world-volume of $Dp$-branes in Type-II string theory. In this more general
case  the CFT/AdS correspondence is really a SYM/SUGRA correspondence,
between the, generically non-conformal 
 $\CN=4$ super-Yang--Mills theories on the $(p+1)$-dimensional world-volumes,  
 and the 
appropriate scaled up throat geometries, as in \refs\rtelaviv.
 We also keep the
original, non blown-up, black brane configurations for further use.

In section 3 the classical free energy is evaluated for the various classical
manifolds without yet taking into account  the neccesity to modify the manifolds
at small scales. The classical dominance
of the manifold $\Xbh$ is obtained. The evaluation of the corresponding
 specific heat offers
a first glimpse at the phenomenon of lack of bulk decoupling for $p\geq 5$.
 In section 4  the
impact of higher derivative
operators coming from integrating out massive string modes in the
bulk of the holographic manifold is considered. Indications are found that the
gauge theory is self-contained and exhibits  no extra stringy scale. In section
5 the Hagedorn
censorship and its limitations are established. This is done by estimating the
modifications needed to reflect the  singular nature of the
vacuum manifolds at the string scale,  
 both in gauge theory and string theory language. In the process,  
situations in which Hagedorn
scales may be exposed are discussed. In particular
we mostly study the SYM/SUGRA correspondence in dimension $(p+1)$ within a
specific
limiting procedure: the large
$N$ expansion of the full string-brane theory is taken first, 
and each term in this
expansion is then expanded in the limit $N\rightarrow \infty$
with $g_s N = {\rm fixed}$,  the supergravity regime
corresponds to $g_s N \gg 1$. For $g_s N \ll 1$ 
one is on much less firm
ground. String corrections are of order one, curvatures are large and the
classical supergravity picture gets fuzzy. Nevertheless, as it is there that
Hagedorn features  surface,  we do venture into that region.

 The SYM/SUGRA correspondence
actually may intrinsically set its own limitations. On the gauge theory
side, the description ceases to be well defined in the absence of more
ultra-violet data for $p\geq 5$, whereas the gravitational side shows unstable   
thermodynamics, with   negative values  of the  
specific heat in the same range of dimensions.
 This would indicate a breakdown of the correspondence.
It turns out that for $p\geq 5$ this is indeed the case, the gravitational
systems have only regions of negative specific heat at the
leading (planar gauge or classical gravitational) order. At the next to leading
order holography breaks down for the same values of $p$ ,  and the 
 extensive
ten-dimensional 
behaviour is restored. This gives an additional point of view from which 
it seems that modifications of the correspondence are imminent for these
large values of $p$ \refs\rsmal.  In section 6
this breakdown of holography for $p\geq 5$, is demonstrated on the basis of the
$1/N$ corrections to the free energy. Having dealt with the Hagedorn
regime from the SYM/SUGRA point of view 
we conclude by studying  that region in section 7,  from a more directly 
stringy point of view, this is done both by applying the analysis of section 5
to background configurations which are not fully blown-up and by studying the 
system now at weak coupling. In
 the first case the Hagedorn transition emerges,  
but it does so on both sides of the correspondence. In the second 
situation $p=5$ is recaptured as a critical dimension and open strings in the 
presence of branes behave thermodynamically not so different than
closed strings.

\newsec{Classical Throat Geometries} 
In this section we discuss the two types of bulk manifolds whose
boundary is ${\bf S}^1 \times {\bf T}^p$, with $p<7$. The cases
$p\geq 7$ cannot be realized in terms of asymptotically flat
brane solutions, and we do not consider them. 
 
\subsec{Vacuum Manifold}

The construction of the vacuum manifold, denoted $\Xvac$,  starts 
 from the closed-string massless backgrounds around the euclidean
 extremal
 $Dp$-brane with toroidal topology ${\bf S}^1_{\beta} \times
 {\bf T}^p_{L}$, \refs\rdp.
In terms of the harmonic function 
\eqn\harmh{
H= 1+ \left( {b \over r}\right)^{7-p} \,, }
we have the following complete geometry, which we shall denote by 
$EDp$: 
\eqn\dpsol{
ds^2 (EDp) = {1\over \sqrt{H}}(d\tau^2 + d{\vec y}^{\ 2}) + \sqrt{H} (dr^2 +
 r^2 d\Omega_{8-p}^2)  \,.}
In addition, we also have profiles for the dilaton and the Ramond--Ramond
(RR) $(8-p)$-form with $N$ units of flux through the angular sphere
${\bf S}^{8-p}$:     
\eqn\dpprof{\eqalign{
e^{-2\phi} =& e^{-2\phi_{\infty}} H^{p-3 \over 2} \cr
F_{r\tau {\vec y}} =& e^{-\phi_{\infty}} F'_{r\tau {\vec y}} =
 e^{-\phi_{\infty}}\, \pt_r {1\over H} \,.}}  
The torus identifications are $\tau \equiv \tau +\beta$ and ${\vec y} \equiv 
{\vec y} +L$. The `charge radius' is defined by 
\eqn\defb{
 b\equiv \ls \, (g_s N)^{1\over 7-p} 
\,.} 
The precise relation between $g_s$ and the asymptotic string coupling is
$$
 g_s = d_p \, (2\pi)^{p-2}  e^{\phi_{\infty}} =
e^{\phi_{\infty}} \, {(2\pi)^{7-p} \over {\rm Vol} ({\bf S}^{8-p})}
.$$
The `gauge theory' scaling takes $\ls \rightarrow 0$ and 
$ r \rightarrow 0$,
 holding
the energy scale  $u= r/\ls^2\,$ fixed. This corresponds, in the microscopic
 D-brane
picture, to the decoupling limit of open-string excited modes, while keeping
fixed the massless Higgs expectation values (i.e. the moduli space) of
the low energy SYM theory.   In the geometry \dpsol, this limit               
blows-up the throat of the extremal $Dp$-brane solution,
 to obtain a new manifold, denoted $\Xvac$, \refs\rmalda, \refs\rtelaviv:  
\eqn\vacpp{ 
{ds^2 (\Xvac) \over \ls^2}  =  {u^{7-p \over 2} \over g_{\rm YM} \sqrt{d_p N}} (d\tau^2 + 
d{\vec y}^{\ 2})  + {g_{\rm YM}\sqrt{d_p N} \over u^{7-p \over 2}} du^2 + 
g_{\rm YM}\sqrt{d_p N} u^{p-3 \over 2} d\Omega_{8-p}^2 \,, }
where the Yang--Mills coupling is defined by  
 \eqn\ymc{g_{\rm YM}\sqrt{d_p N} = \ls^{p-3 \over 2} \,\sqrt{g_s N}.
}
It should be noted that only the fully blown-up manifolds are known 
for the case of CFT on a sphere, as in \refs\rwituno, \refs\rwitdos. A
brane realization incorporating the spherical geometry of the world-volume,
and allowing a smooth transition into flat ten- or eleven-dimensional
Minkowski space-time, is not yet available.  
 
Restoring  
the radial variable $r=u\,\ls^2$, this limit is  the near-horizon
scaling, consisting of neglecting the $1$ in the harmonic function $H$:  
\eqn\vacp{
ds^2 (\Xvac) = {r^2 \over b_r^2}\, (d\tau^2 + d{\vec y}^{\ 2} )
 + {b_r^2 \over r^2} \, dr^2
+ b_r^2 \,d\Omega_{8-p}^2 \,,}
where we have defined the  quantity:
\eqn\defbr{ 
b_r^2 \equiv r^{p-3 \over 2} b^{7-p \over 2}. } 
In this way the metric is cast in the form of a `product' $AdS_{p+2} \times 
{\bf S}^{8-p}$, both factors
 with a varying radius of curvature, equal to $b_r$. Thus 
the $\alpha'$ expansion parameter, i.e. the scalar curvature in string units
is given by
\eqn\curv{ \alpha' R \sim \left({\ls \over b_r }\right)^2       
\,.}
For $p>3$ the curvature is small as $r\rightarrow \infty$ and large as
$r\rightarrow 0$, and the other way around for $p<3$. The critical point
where the curvature becomes of order one in string units is at 
\eqn\critcurv{
r_c \sim \ls\, (g_s N )^{1\over 3-p} \sim b\, (g_s N)^{4\over
(7-p)(3-p) } 
.}

The supergravity  description (closed strings propagating in a smooth,
`large' background)  will be good at large coordinate radius  
for $p>3$, and near the horizon for $p<3$. 
 In order for the smooth region
to be large in string units we require $g_s N \gg 1$. 
As pointed out in \refs\rtelaviv, this is in nice agreement with the
idea that the radial coordinate represents (via the Higgs expectation
values of a probe $D$-brane)
 the renormalization group scale in the gauge theory language.
In this dictionary, the ultraviolet region is the upper part of the
throat, and the infrared is represented by the near-horizon region. Thus,
the breakdown of the supergravity description in the ultraviolet for
$p<3$ is in correspondence with the asymptotic freedom (i.e. perturbative
nature) of SYM at short distances, in 
less than four dimensions. Conversely, the breakdown of the supergravity
description in the infrared for $p>3$ agrees with the non-renormalizability
of SYM above four dimensions; i.e. these theories flow to free (perturbative)
fixed points in the infrared. In other words, there is a complementarity
between the SYM and SUGRA descriptions, with a cross-over roughly
located at $r\sim r_c$, where the geometry becomes `fuzzy'.   

There is also a non-perturbative threshold for $p\neq 3$, due to the fact
that the dilaton has a non-trivial profile. At a radial coordinate of 
order
\eqn\dilr{ r_{g} \sim r_c \,  N^{4\over (7-p)(p-3)},}
the throat develops local strong coupling $e^{\phi (r_g)} \sim \CO(1)$. At 
these points, lying in the ultraviolet (infrared) for $p > (<) 3$, we must
go over dual descriptions, as pointed out in \refs\rtelaviv. In the present
paper, we take the large $N$ limit before any other expansion,  
 so that  $N$ is larger than any dimensionless parameter
formed out of couplings or ratios of scales in the problem. In particular,  
we take $N\gg g_s N$, and then we can always postpone the non-perturbative
threshold, as compared to other thresholds which depend only on the
combination $g_s N$.

 One such threshold that will interest us in this paper is the onset of
$\alpha'$ corrections due to light states with support in the background
geometry under consideration. In particular, light string 
winding modes will set in when
the proper length of the non-contractible circles is of order the
string scale. For termal winding modes this happens at radial coodinates  
\eqn\rhag{
r_{\beta } = b \left( {\ls \over \beta}\right)^{4\over 7-p} = {b\over          
(g_s N)^{4\over (7-p)^2}} \,\left({b\over \beta}\right)^{4\over 7-p}, }
and we also have a  threshold for spatial winding modes, at $r_L$ defined by
substituting $\beta$ by $L$ in this equation. Below these radii, the
semiclassical propagation of Type-II strings is weighted in perturbation
theory by the effective expansion parameter $(g_s / m \ls )^2$ for
each closed string loop. The mass $m < 1/\ls$ corresponds to
the new light winding modes. This means that string perturbation theory
breaks down, even if we are in a region where the nominal value of the local
string coupling is  small, $e^{\phi (r)} \sim N^{-1}$, 
 for large, but finite,  $N \gg g_s N \gg 1$.

In the following, it will be important to distinguish between the
blown-up throat manifold \vacpp\ or \vacp, denoted $\Xvac$, and the
complete $Dp$-brane geometry in \dpsol, denoted $EDp$. 
The meaning of the radial coordinates $r>b$ in
$\Xvac$ is associated to the ultraviolet behaviour of the system
in the field theory limit \vacpp, which involves, in particular,  
the decoupling of the string oscillators. There are, however, indications
that (\refs\rigg, \refs\rdalw) keeping the end of throat or `neck' at
$r\sim b$, is associated to the stringy thresholds in the D-brane sector. 
In the following, we will make explicit this distinction between the
full system, with the stringy thresholds kept in place, and the truncated
system: SYM at weak coupling, and (infinite) throat geometry at strong
coupling.

A crucial difference between these
throat geometries with flat boundary space-time and the time slicing  
adequate for finite spherical topology in \refs\rwituno, is the occurrence
of arbitrarily small non-contractible loops down the throat. This means
that, for example, Hagedorn scales  occur down the throat of $\Xvac$
no matter how small we dial the ratio $L/\beta$ at the boundary.

\subsec{Black-Hole Manifold}

The $\Xbh$ or black-hole-type manifold, with the same asymptotics
${\bf S}^1_{\beta} \times {\bf T}_{L}^p$ is obtained from the blow-up of the
throat of the near-extremal black $Dp$-brane. We denote this manifold by
$BDp$,  with metric \refs\rdp 
\eqn\bbrane{ds^2 (BDp) =
 {1\over \sqrt{H}}\left(h\,d\tau^2 + d{\vec y}^{\ 2}\right) + \sqrt{H} 
\left({dr^2
\over h} +
 r^2 d\Omega_{8-p}^2 \right),  }  
where we have introduced the Schwarzschild-like harmonic function:
\eqn\hache{ h= 1-\left({r_0 \over r}\right)^{7-p}, }
and the function $H$ also depends on $r_0$: 
\eqn\hachem{\eqalign{ H &= 1+ \left({r_p \over r}\right)^{7-p} \cr
r_p^{7-p} &= -{r_0^{7-p} \over 2} + \left( b^{2(7-p)} + 
{r_0^{2(7-p)} \over 4} \right)^{1/2}.}} 
The dilaton and Ramond--Ramond (RR)  backgrounds have the same form as in \dpsol, with
$H$ defined in \hachem, so that the extremal solution   is the limit
$r_0 \rightarrow 0$ of the black-brane solution \bbrane.  
The extremality parameter $r_0$ gives the location of the horizon. In
the euclidean section, spacetime is restricted to $r\geq r_0$.

The gauge theory regime is defined by the near-horizon limit, i.e. by
the restriction  $r, r_0 \ll b$, and we obtain
the scaled-up throat, or   $\Xbh$ manifold as
\eqn\bhp{
ds^2 (\Xbh) = {r^2 \over b_r^2} \left( (1-r_0^{7-p} /r^{7-p} ) d\tau^2 + d{\vec
y}^{\ 2} \right) + {b_r^2 \over r^2} {dr^2 \over (1-r_0^{7-p} /r^{7-p} )} +
b_r^2 d\Omega_{8-p}^2 \,.}

\ifigc\legend{Inverse Hawking temperature as a function of $r_0$  
for the  $p<5$ manifolds.} 
{\epsfxsize3.50in\epsfbox{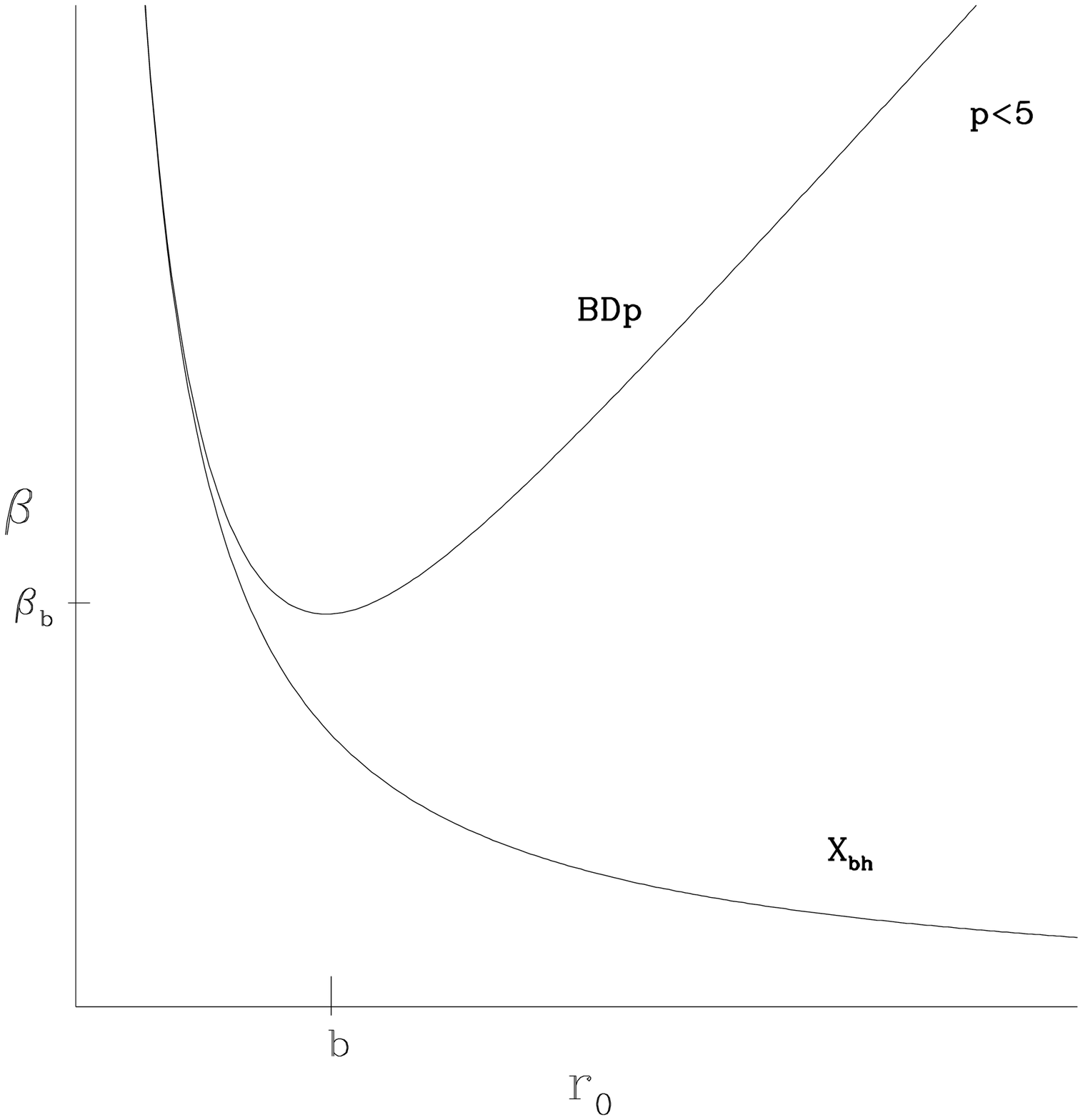}}

The euclidean section \hachem, or \bhp\  is smooth at $r=r_0$ only if the
variable
\eqn\smooth{
\theta = {\tau \over 2} \sqrt{\pt_r g_{\tau\tau} \over \pt_r g_{rr}}}
has period $2\pi$. So, if the period of $\tau$ is $\beta$, we obtain  
\eqn\eqbeta{
\beta = 4\pi \sqrt{{\pt_r g_{rr} (r_0) \over \pt_r g_{\tau\tau} (r_0)}}
 = {4\pi \over 7-p} \,r_0 \sqrt{H(r_0)}. }
This equation defines the canonical thermal ensemble, providing the
functional dependence of the mass as a function of temperature.
It applies not only to the scaled up throat manifold $\Xbh$ in \bhp, but also
to the complete black-brane geometry,  $BDp$ in \bbrane. 

The behaviour is rather different from the case of spherical
boundary, studied in \refs\rwitdos, where the black-hole manifold
$\Xbh$ could only be defined for sufficiently large temperature:
$\beta_0 < b$, with $\beta_0^{-1}$ the temperature at the origin
of the AdS space. For the range of allowed temperatures, there
was a branch of large black holes with radius $r_0 > b$ and
positive specific heat, along with a brach of small ones $r_0 <b$,
and negative specific heat. The canonical ensemble was dominated 
at high temperature $\beta_0 < b$ by
the large black holes with positive specific heat. 

\ifigc\legend{Inverse Hawking temperature as a function of $r_0$  
 for  the  $p=5$ manifolds.}
{\epsfxsize3.50in\epsfbox{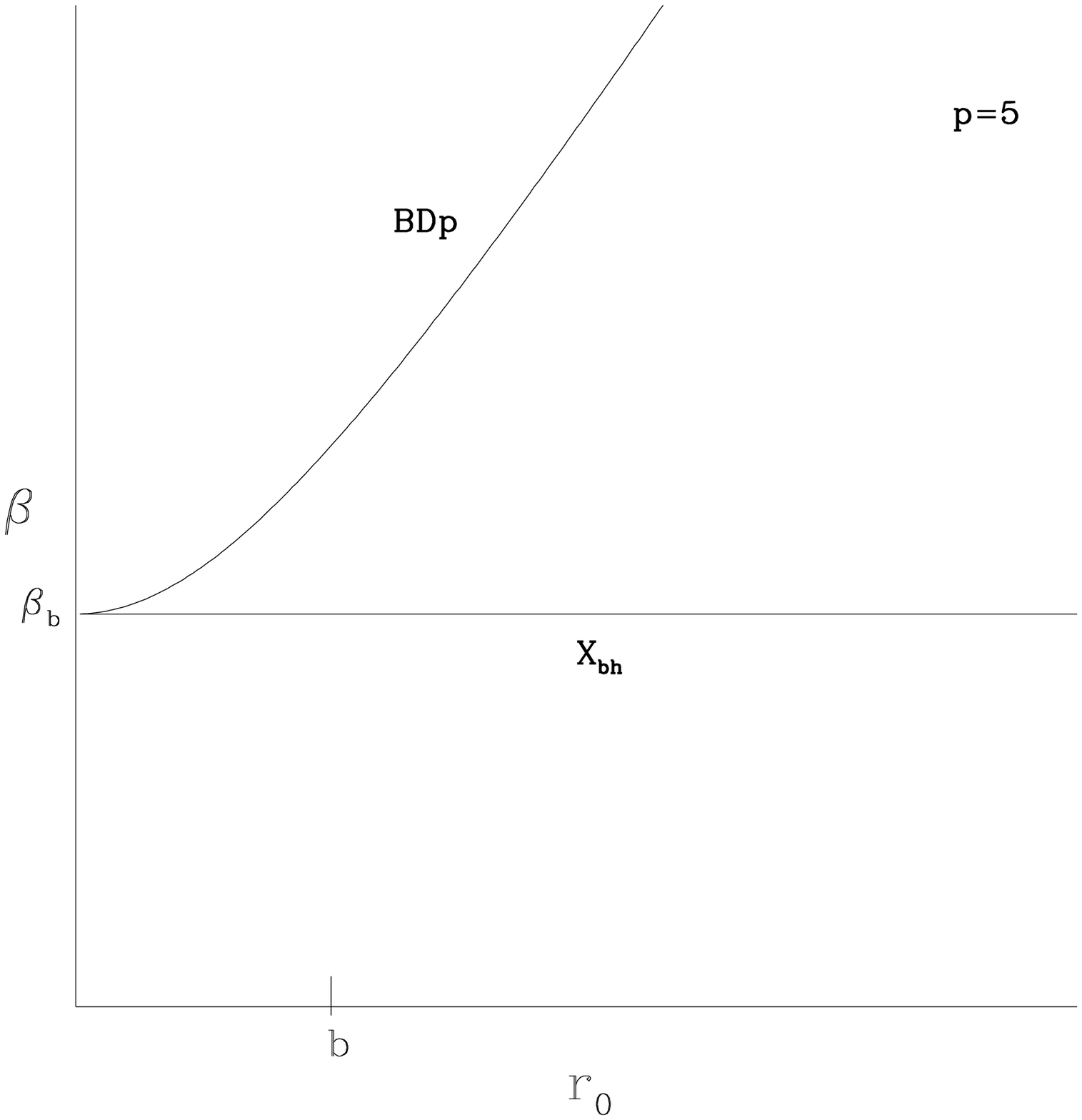}}

For most values of $p$, the situation is exactly opposite in the
case of the $BDp$ metric. For $p<5$, the complete
 curve \eqbeta\ has
 a minimum at radius $r_0 \sim b$. It grows without bound 
in the near-extremal or gauge theory 
 regime $r_0 \ll b$, with asymptotics 
$
\beta (r_0 \ll b)  \sim b (b/r_0)^{5-p\over 2}
$. 
This is precisely the temperature-horizon relation for the scaled
up $\Xbh$ manifold:
\eqn\eqbexa{ \beta (\Xbh) ={4\pi \over 7-p}\,  b \,\left({b\over r_0}
\right)^{5-p\over 2} .} 
This is a branch with positive specific heat, the one that can be
related to gauge theory thermodynamics. In the complete $BDp$ metric,
 there is also the  usual Schwarzschild
regime, $r_0 \gg b$, with the law (universal for all values of $p$):  
\eqn\sxh{\beta(r_0 \gg b) \sim r_0 ,} 
 and negative specific heat. We shall see that this
 branch (which is absent in $\Xbh$) is thermodynamically unfavoured in
the canonical ensemble. Thus,  $BDp (p<5)$ can only be defined
for sufficiently low temperatures $T< T_b \sim b^{-1}$, and the branch with
positive specific heat $r_0 \ll b$, is defined for arbitrarily low temperatures,
unlike the case of spherical boundary, (see fig. 1.)              

For $p=5$, a marginal case, $\beta (r_0)$ is a monotonously increasing
function with vanishing derivative at $r_0 =0$. The metric $BDp (p=5)$ is
defined only for temperatures $T < T_b \sim b^{-1}$, and has negative specific
heat for all temperatures. The scaled-up throat $\Xbh$ has constant 
temperature, independent of the value of $r_0$, $T= 1/2\pi b$, 
 and thus formally
has infinite specific heat (fig. 2.)          

 The last case, $p=6$, shows a similar behaviour. There is a single
regime of the complete black-brane metric $BDp (p=6)$ 
with negative specific heat, but now defined for arbitrary
high temperatures. The scaled-up throat $\Xbh (p=6)$ has also negative
specific heat and is only defined for very high temperatures, with
 asymptotics
given by eq. \eqbexa, which still applies (fig. 3.)

 We see that it is unlikely that $\Xbh$ is related
to gauge theory thermodynamics in the cases $p=5,6$, since they lack
a phase with positive specific heat at low temperatures.

 \ifigc\legend{Inverse Hawking temperature as a function of $r_0$  
 for the   $p=6$ manifolds.}  
{\epsfxsize3.50in\epsfbox{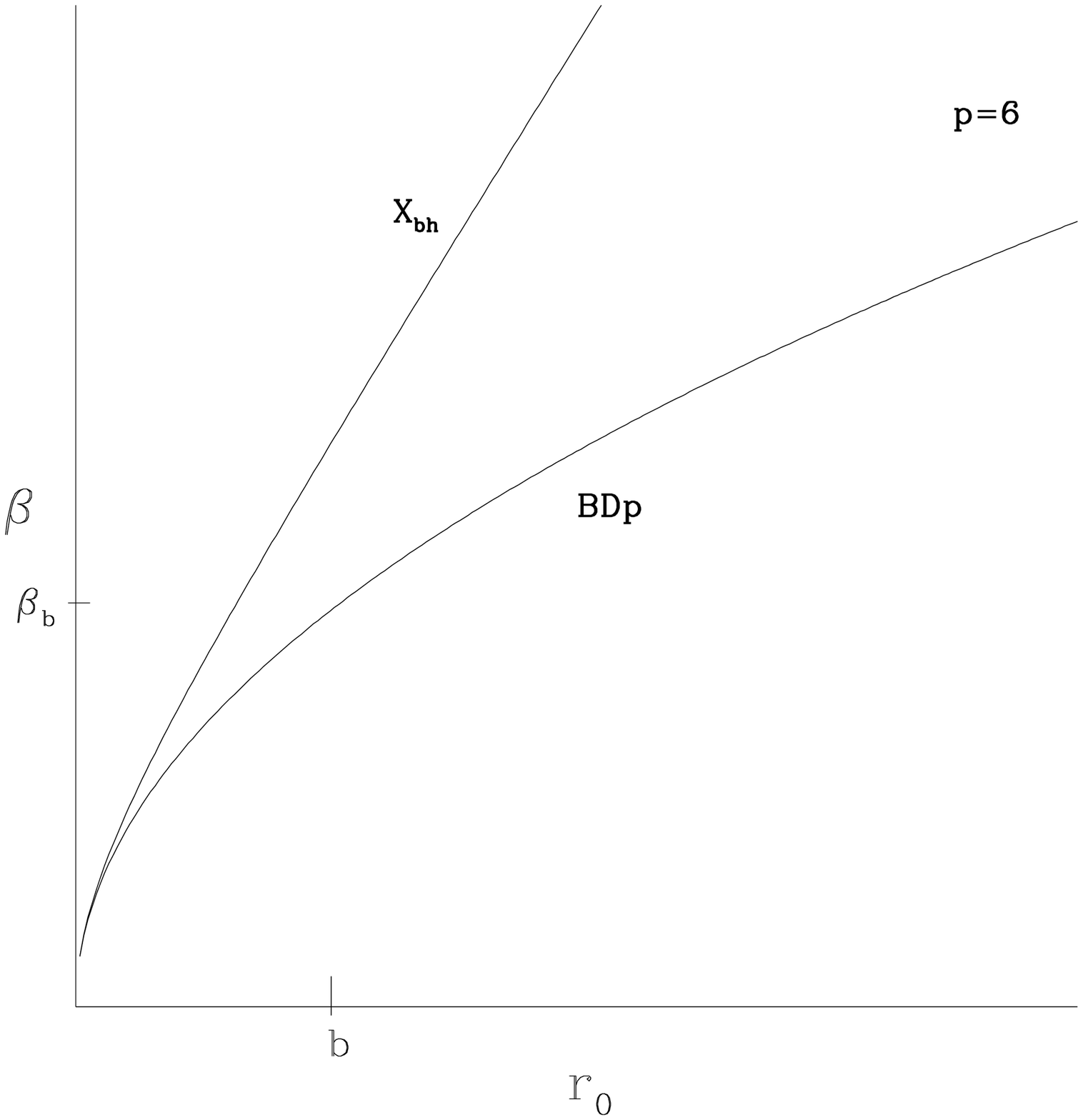}}

The local red-shifted temperature in $\Xbh$ diverges at the horizon.
This is a general property of any horizon, and it would lead us
to suspect that a Hagedorn regime is reached when this local
red-shifted temperature is of order the string scale. On the other
hand, the topology of the thermal modes in $\Xbh$ is trivial. What happens
is that the `Hagedorn' region at the tip of $\Xbh$ is defined by a
disc with a proper radius of the order of the string length. Therefore,
a thermal winding mode in the local Hagedorn region can unwind at no  
cost of energy. Indeed, string `propagation' is completely smooth 
around the horizon of the euclidean $\Xbh$ manifold. Hagedorn effects
should be taken into account when {\it metastable} light winding modes
can be identified. This would happen for example if the local temperature
in $\Xbh$ reaches the string scale at large radial distance from the
horizon, so that it costs a large amount of energy to `unwind' the
thermal `winding' modes. In this case the geometry of $\Xbh$ is that
of a very `thin cigar' and this can be arranged only when the asymptotic
temperature itself is of the order of the string scale.

\newsec{Classical Free Energy and Black-Hole Dominance}  
For a boundary of topology ${\bf S}^1 \times {\bf S}^3$, the bulk
gravitational theory exhibited a phase transition at some critical
temperature \refs\rwituno, \refs\rwitdos. Here we repeat these considerations
for the case at hand.    
The analysis in this section is based on
 the classical supergravity approximation,
as in \refs\rkt,   
although we have argued in section 1 that it should become invalid  sufficiently
far down the throat of  $\Xvac$.       Such effects will be analyzed separately
in section 5.  

The action, (or planar free energy in the gauge theory interpretation), of
any of the  manifolds under consideration is
\eqn\fren{ I(X)_{c\ell} = \beta M(X) - S(X). }
We shall consider first the cases of the general metrics, $EDp$ and $ BDp$,
given by  
\dpsol\ and \bbrane, respectively, and then recover the actions
of the scaled throat geometries $\Xvac$ and $\Xbh$  by taking 
appropriate  near-horizon  limits.

In formula \fren,  $M$ is the ADM  mass,   derived from the asymptotics of the
$d=10-p$ black hole obtained upon dimensional reduction in the 
wrapped ${\vec y}$-directions \refs\radm: 
\eqn\adm{g_{\tau\tau} \sim 1- {16\pi G_d M \over
 (d-2){\rm Vol}({\bf S}^{d-2} ) r^{d-3}},}  
with $G_d = G/L^p$ the $d=10-p$ dimensional Newton constant, and 
$G \equiv G_{d=10} = 8\pi^6 \alpha'^4 e^{2\phi_{\infty}}$. The
$d$-dimensional metric is obtained by dimensional reduction of \hachem\
and subsequent rescaling to the Einstein frame: 
\eqn\eins{ ds^2 \rightarrow e^{-4(\phi_d -\phi_{\infty}) \over d-2} ds^2 , }
with $\phi_d$ the $d$-dimensional dilaton:
\eqn\dild{e^{-2\phi_d} = e^{-2\phi} \sqrt{{\rm det}(g_{{\vec y}{\vec y}})} . } 
The resulting Einstein-frame black-hole metric is
\eqn\bhe{ds^2_d = H^{3-d \over d-2} \,h\, d\tau^2 + H^{1\over d-2} (dr^2 /h +
r^2 d\Omega_{d-2}^2 ), }  
and matching the asymptotics to \adm\ one obtains
\eqn\mass{M (BDp) 
= {{\rm Vol}({\bf S}^{8-p}) L^p \over 16\pi G} \left( (8-p) r_0^{7-p} + 
(7-p) r_p^{7-p} \right). } 

There are two limiting regimes: the Schwarzschild regime, where
the horizon radius is much larger than the charge radius $r_0 \gg b$,  
\eqn\masssch{M(r_0 \gg b)_{BDp}
 \sim (8-p) \,{\Vol L^p \over 16\pi G}\, r_0^{7-p}\,,} 
and the opposite, near-extremal regime:
\eqn\massex{ M(r_0 \ll b)_{BDp} \sim M_{EDp} + {9-p \over 2}
  \,{\Vol L^p \over 16\pi G}\, r_0^{7-p} \,,} 
with 
\eqn\exms{M_{EDp} = (7-p)  \,{\Vol L^p \over 16\pi G}\,
 b^{7-p}}
 the mass of the extremal $Dp$-brane
system.  
The corresponding `mass' of the `throats' $\Xvac$ and $\Xbh$ is   
defined by these near-extremal limits
\eqn\masth{
\eqalign{M(\Xvac) =& M_{EDp} \cr   
M(\Xbh) =& M_{EDp} +{9-p \over 2}
  \,{\Vol L^p \over 16\pi G}\, r_0^{7-p} \,.}}

The entropy can be obtained in the standard fashion from \bhe:
\eqn\entr{ S = {A_H \over 4G_d} = {\Vol L^p \over 4G} r_0^{8-p} \sqrt{H(
r_0)}  
}
or, using the temperature formula \eqbeta:
\eqn\bent{ S= (7-p) \,{\Vol L^p \over 16\pi G}\, \beta \, r_0^{7-p} \,. } 
Written in this way, this formula for the entropy is exact and applies
to all cases $BDp$ and $\Xbh$. The entropy of the `vacuum' manifolds
$EDp$ and $\Xvac$ is obtained as the $r_0 \rightarrow 0$ limit of this
expression and it vanishes identically.

By definition, the thermodynamics of the manifolds $\Xvac, \Xbh$ is equivalent
to the one of $EDp, BDp$ in the gauge theory regime, i.e. for temperatures
$T\ll T_b \sim b^{-1}$. We first focus on the analysis of the field-theory
limit thermodynamics and postpone to section 7 the discussion of the other
Schwarzschild-like regimes of the $BDp, EDp$ pairs.

The free energy of $\Xbh$, obtained through \fren, is 
\eqn\ac{I(\Xbh)_{c\ell} = - {5-p \over 2}  \,{\Vol L^p \over 16\pi G}\,
 \beta \, r_0^{7-p} + 
\beta M_{EDp} \,.} 
The action of $\Xvac$ follows by subtituting $r_0 =0$ in this
formula. One finds that it is completely given by a vacuum energy\foot{By
using in \fren\ the ADM definition of mass to calculate the classical action,   
we bypass subtleties in the treatment of boundary terms at infinity, as
in \refs\rwitdos. Such subtleties arise if we evaluate directly the canonical
action on the solution, and amount to the correct prescription of \fren, with
$M$ the ADM mass, and $S$ the Bekenstein--Hawking entropy \entr.} 
\eqn\vacu{I(\Xvac)_{c\ell} = \beta M_{EDp}.} 
Therefore, the action difference between $\Xvac$ and $\Xbh$, or the planar
free energies,   becomes
\eqn\diff{ \beta F(\beta)_{\rm planar} =  I(\Xbh)_{c\ell} - I(\Xvac)_{c\ell} =  
- N^2 \,(5-p)\, C_0 \, (g_s N)^{2(p-3) \over 7-p} \, L^p \, b^{\gamma -p} \, 
\beta^{-\gamma} \,, }  
with $C_0$ the positive   constant,
\eqn\eqc{
C_0 = {(2\pi)^{7-2p} \over 2\,{\rm Vol}({\bf S}^{8-p})} \,\left({4\pi\over 
7-p}\right)^{2(7-p)\over 5-p} \,,}
  and $\gamma$ the
critical exponent 
\eqn\gammpl{ \gamma = {9-p \over 5-p}.} 
This critical exponent 
 is larger than that of
free fields $\gamma_{\rm free} = p$, for all $p$ except the conformal
one $p=3$ where  $\gamma = \gamma_{\rm free}$. 

The thermodynamical balance in \diff\ yields $\Xbh  $ dominance for
$p<5$  at all temperatures.  
This contrasts with the behaviour with spherical boundary topology,
\refs\rwituno, \refs\rwitdos, where $\Xvac$ dominated the low-temperature
phase. Thus, within this approximation, there is no large $N$ `flop' 
between $\Xvac$ and $\Xbh$, as long as  
as $p<5$.

 For $p=5$ we have a marginal behaviour, with exactly
balanced free energies between $\Xvac$ and $\Xbh$ at the temperature at
which $\Xbh$ is defined, $T(\Xbh) = 1/2\pi b$. However, for any other
temperature, only $\Xvac$ contributes to the canonical ensemble. Thus,
generically, planar thermodynamics is dominated by vacuum energy for $p=5$.

The situation is analogous for $p=6$. In this case, the vacuum-subtracted
 free energy of
$\Xbh$ is {\it positive},  and therefore thermodynamically suppressed. 
 $\Xvac$ dominates at all temperatures.     

This concludes our discussion of the purely classical thermodynamical  
properties of the blown-up manifolds $\Xvac$ and $\Xbh$. We find a neat
difference in the canonical ensemble, depending on whether $p$ is larger
or smaller than five but, for a given dimension, we {\it do not} find
${\bf T}^p$ analogues
 of the finite-volume phase transition reported in \refs\rwituno,  
\refs\rwitdos\ for the spherical case.    In the rest of the paper
we investigate to what extent these conclusions are modified by  
stringy effects, either
 in the bulk of the $X$-manifolds, or perhaps localized 
around singular loci.

\newsec{Short-Distance Corrections}
According to the dictionary developed in \refs\rmalda, the $\alpha'$ or
low-energy expansion of the string theory in the AdS space corresponds
to the strong-coupling expansion of the large $N$ gauge theory in 
inverse powers of the 't Hooft coupling: $\alpha' /b^2 = 1/\sqrt{g_{\rm 
YM}^2 N}$. 
On the other hand, string-loop corrections yield the topological expansion
of the gauge theory diagrams: $g_s^2 = (g_{\rm YM}^2 N /N)^2$, and
semiclassical Yang--Mills instanton corrections are related to
D-instanton processes in the Type-IIB theory \refs\rgreenbk, \refs\rkinst,
\refs\rchin.  
In the low-energy approximation, 
the effect of the massive string states propagating in the throat geometries
$\Xvac, \Xbh$,  can be incorporated into the
low-energy supergravity action by means of higher-derivative operators
in the massless fields, suppressed by powers of the string length squared
$\alpha' =\ls^2$. The form of the effective action induced within string
perturbation theory is
\eqn\indeff{
I_{\rm eff} \sim {1\over g_s^2} \,
\sum_{g\geq 0} \sum_{n\geq 1} (g_s^2)^g \, (\alpha')^n  
 \,\int   {d{\rm Vol} \over \alpha'^5} \, \CO_{(g,n)} \,.} 
If only the fields in \dpsol\ and \dpprof\ are excited, the important
operators of engineering dimensions ${\rm dim}(\CO) = 2n = 2(n_c +
n_{\phi} + n_f)$,  generated at $g$-loop order,  take the form
\eqn\gencorr{
\CO_{(g,n_i)} \sim  \left(e^{-2(\phi-\phi_{\infty})}\right)^{1-g}
\, R^{n_c} \, (\pt\phi)^{2n_{\phi}} \, |F_{p+2}'|^{2n_f} 
, }
where $R$ stands for the Riemann tensor or some contraction thereof. 
The classical values $\bra \CO_{(g,n)}\ket_X$ on the manifold $X$ determine 
the leading corrections in first-order perturbation theory in the
coefficient of such operators. A proper treatment of the $g_s$ and
$\alpha'$ expansion to $k$-th order should also include the effect
of the operators \gencorr, through the back-reaction on
 the manifold geometry, to order $k-1$. 
However, for the purposes of the qualitative dependence on the various
dimensionful parameters, like for example the temperature, we can
estimate the effects of \indeff\ by its expectation value on the
large $N$ uncorrected saddle point $I_{\rm eff} \sim \sum \int
\bra \CO_{\rm eff} \ket_X$, provided we expand all expressions in
power series in $g_s$ and $\alpha'$.

Under these circumstances, we can
 estimate the effects of these short-distance corrections through
scaling arguments (for a precise calculation of a leading correction
in the $p=3$ case, see \refs\rgkts.) 
We consider 
 the black-hole manifold $\Xbh$, 
and recover $\Xvac$ as the $r_0 \rightarrow 0$ limit. 
It is useful to introduce yet another system of coordinates:
\eqn\rhodef{
z^2 \equiv b_r^2 = r^{p-3 \over 2} b^{7-p \over 2}, }
so that, for $p\neq 3$,  $r^2 /b_r^2 =
 z^{\alpha} /b^{\alpha}$ with $\alpha = 2(7-p)/(p-3)$.
 Define also
\eqn\tdef{
(t,{\vec x}) \equiv b^{-\alpha/2} (\tau, {\vec y}), }
thereby casting the near-extremal metric \bhp\  in the form
\eqn\rhome{
ds^2 (\Xbh) = z^{\alpha} \left( (1-z_0^{2\alpha} /z^{2\alpha} )
dt^2 + d{\vec x}^{\ 2} \right) + {C dz^2 \over (1-z_0^{2\alpha} /
z^{2\alpha} )} + z^2 d\Omega_{8-p}^2 \,,}
with $C= 16/(p-3)^2$. This representation, without explicit dependence
on $b$ or $\alpha'$, is useful for scaling arguments, because a local
invariant of dimension $D$ takes the form $z^{-D} \, f(z_0^{2\alpha}
/ z^{2\alpha} )$. For example, $D=2$ for the curvature scalar $R$ or
the dilaton gradient squared $(\partial \phi)^2$.
In addition, translational invariance with respect to the $(t,{\vec x})$
coordinates ensures  the effective action is proportional to
$\beta_t L_x^p = b^{-\alpha (p+1)/2} \beta L^p$.
 
 We then find the following
near-horizon asymptotics for the relevant invariants contributing to
the effective action:
\eqn\dilsca{\eqalign{ e^{-2\phi} \sim & {1\over g_s^2} \left( {b\over z}
\right)^{2(7-p)} \cr
|F_{p+2} |^2 \sim & {1\over z^2} e^{-2\phi} \cr
R\sim & {1\over z^2}\,.}  }
 
The overall scaling of each operator in \gencorr\
 depends only on the total $\alpha'$ counting index
$n=n_c + n_{\phi} + n_f$.
Thus, a general operator of dimension $2n$, generated at $g$ loop order,
 has the following scaling when evaluated on the
solution \rhome:
\eqn\scalr{
\bra \CO_{g,n} \ket_{\Xbh}  \sim N^{2-2g}  \, {(g_s N)^{10-2n\over 7-p}
\over (g_s N)^{2-2g}}\, \beta \,L^p \,b^B\, \left[ z^A
 f(z^{2\alpha}/z_0^{2\alpha}) \right]_{z_-}^{z_+} \,,}
where $z_{\pm}$ denote the limits between which the radial coordinate is
integrated. The power of $b$ is
\eqn\pob{B=2n-10 +2(7-p)(1-g) - {\alpha \over 2} (p+1), }
and $A$ is fixed accordingly by dimensional analysis. Restoring the $r$ variable
through \rhodef, we find the following result for the effective action:
\eqn\cor{
I_{\rm eff} = \beta E_{\rm vac} + N^2 \sum_{g=0}^{\infty} \sum_{n=1}^{\infty}
\, I_{g,n}  \, (g_s N)^{2-2n\over 7-p} \, \left({g_s
 N \over N}\right)^{
2g} ,}
where
\eqn\genterm{I_{g,n} = -C_{g,n} \, (g_s N)^{2(p-3)\over 7-p} \,
L^p \, b^{\gamma_{g,n} -p} \, \beta^{-\gamma_{g,n}}\,, }
and the critical exponent is:
\eqn\critex{ \gamma_{g,n} = {9-p \over 5-p} - {p-3 \over 5-p} \left( n-1-
g(7-p)\right) = {(3-p)n +g(10p -p^2 -21) + 6 \over 5-p}. }
The classical term is obtained for $n=1$ and $g=0$.
The estimate of the short-distance corrections to the 
$\Xvac$ free energy
can be done along the same lines, putting $r_0 =0$ in the solution,
and evaluating \scalr\ between two limiting radii $r_{\pm}$. The result
is clearly proportional to $\beta$, and we can naturally interpret it
as a correction to the vacuum energy.

A very interesting feature of \cor\ and \genterm\ is the possibility of
eliminating the separate dependence on $g_s N$ and $b$. One can combine  
the dependence on the effective D-brane string coupling and the 
`transmuted' string scale, $b$, into the single
 dimensionful 't Hooft coupling
\ymc: 
\eqn\gtfe{
I_{\rm eff} = \beta E_{\rm vac} - \sum_{g=0}^{\infty} \sum_{n=1}^{\infty} 
\, N^{2-2g} \,B_{g,n}\,
 (g^2_{\rm YM} N)^{\gamma_{g,n} -p \over p-3} \, L^p \,\beta^{-
\gamma_{g,n}} \,.} 
This result shows  how, at least within the low-energy expansion,
the effects of closed-string excited states have a gauge theory interpretation,
i.e. the only intrinsic scale of the system is the dimensionful Yang--Mills
coupling $g^2_{\rm YM} N$, and the type-IIB string in the blown-up $X$-manifold
`behaves' like a genuine `QCD-string'. This seems to be a non-trivial 
property of the fully blown-up throats $\Xvac$ and $\Xbh$. We shall argue
in section 7 that the transmutation of the $\alpha'$ dependence into
gauge theory parameters is not complete for the non-blown-up manifolds
$EDp$ and $BDp$.           
 
The scaling laws in \gtfe\ admit some generalizations.
For example, one may turn on boundary values for other fields in the
supergravity backgrounds, corresponding to turning on more operators
in the gauge theory. Special care is needed then for the RR potentials,
which have $\CO(N^0)$ scaling in the large $N$ limit. One such example would
be the leading 
large $N$ contribution to the theta angle dependence in the four-dimensional
theory considered in \refs\rwitot. Such contributions to the effective action
scale like a $g=1, n=1$ term, even if they should be considered as `planar'
from the point of view of the 't Hooft expansion.   
 
 The scaling
law for the conformal case $p=3$ can be obtained by direct subtitution  of 
$z=b$ in \dilsca. One gets the same answer by directly plugging
$p=3$ in the final formula \critex.  
The index $\gamma$
does not receive any loop or $\alpha'$ corrections, i.e. $
\gamma_{g,n} = \gamma_{0,1} = 3$.
 
An interesting remark concerns the large order behaviour of $\gamma_{g,n}$.
We find that, at fixed $g$, $$\lim_{n\to\infty}
\gamma_{g,n} \rightarrow +\infty \;\; {\rm if} \;\;p<3 \;\;{\rm or} \;\;p>5 
.$$
Since $\gamma$ is a measure of the effective dimension for massless field
excitations in the large temperature regime, $\gamma_{g,n} \rightarrow\infty$
can be interpreted as a bad high temperature behaviour, in the sense that
the theory exhibits a host of new degrees of freedom at high energies.
This is compatible with the fact that for $p<3$, the supergravity description
becomes inappropriate in the ultraviolet, where D-brane perturbation theory takes
over. The pathology at high temperature for $p>5$ is to be expected in the
light of our previous discussion about the lack of gravity decoupling
for $p\geq 5$.
We also find
$$
\lim_{n\to\infty} \gamma_{g,n} = -\infty \;\;{\rm if}\;3<p<5
.$$
This corresponds to the fact that the ultraviolet is well described by the supergravity
in this range of brane dimensions, and therefore high order corrections in
$\alpha'$ are less and less singular in the high-temperature limit.
 
On the other hand, the large-order behaviour with respect to string-loop
corrections works the other way around. At fixed $n$, $\gamma\rightarrow
-\infty$ for $p\neq 4$. But $\gamma\rightarrow +\infty$ for $p=4$.
The interpretation of this could be that, precisely for $p=4$, the local
value of the string coupling grows up the throat, towards the ultraviolet.

\newsec{Effects of Light Winding Modes} 
Having dealt in the previous section  with short-distance effects coming 
from the massive closed string states propagating in the bulk of
$\Xvac, \Xbh$, we now turn to other limitations of the supergravity description.
Namely, with branes of topology 
 ${\bf S}^1 \times {\bf T}^p$, both $\Xvac$ and $\Xbh$
support non-contractible cycles. In particular, $r=0$ is singular for
$\Xvac$ because these non-contractible cycles shrink to zero size. As pointed
out in section 2, at radial coordinates of order $r_{\beta}$, temporal cycles
 have string length, while spatial cycles do so at the point
\eqn\rhprima
{r_L  = b \left( {\ls \over L}\right)^{4\over 7-p} = {b\over
(g_s N)^{4\over (7-p)^2}} \,\left({b\over L}\right)^{4\over 7-p}. }
 Accordingly, we expect important modifications of the geometry at
these  scales. At high temperatures from the point of view of the
gauge theory, $\beta \ll L$, the corrections to the geometry 
implied by the light thermal winding modes dominate, since
$r_{\beta} \gg r_L$ in this regime. 
 On the other hand, at low temperatures
$\beta \gg L$, effects due to light spatial winding modes dominate.

\subsec{Spatial Winding Modes and Finite-Size Effects} 

In the low temperature regime, $\beta\gg L$, the approximate  supersymmetry
allows us to determine the effective geometry at radial distances
$r\ll r_L$. In this region, spatial winding modes on the ${\bf T}^p$ torus are
lighter than momentum modes. In we want to continue using the supergravity
language,  we can simply perform a T-duality transformation on the
small spatial circles. This produces a metric of $N$ coincident
 $D0$-branes of 
Type-IIA string theory,  
`smeared' over the torus ${\bf T}^p$, which now has a local size
larger than the string scale:
\eqn\smear{
ds^2 ({\widetilde\Xbh}) = {h\over \sqrt{H}}\,d\tau^2 +
\sqrt{H}\,\left(d{\vec{\tilde y}}^{\ 2} + {dr^2 \over h} + r^2 \,d\Omega_{
8-p}^2 \right),  }  
 where we have defined the dual coordinates on the T-dual torus with
identification ${\tilde y} \equiv {\tilde y} + {\widetilde L}$, with
${\widetilde L} = (2\pi)^2 \ell_s^2 /L$, and the harmonic functions
$H$ and $h$ as in \harmh, \hache. At $r\sim r_L$,
 this manifold is glued to the previous
one \bbrane\ or \bhp\ by a `neck' of stringy size and $\CO(1)$ curvatures.
The cross-over temperature at which we go over the T-dual language is
determined by $r_0 \sim r_L$:
\eqn\fst{
T(X_{\rm bh}\leftrightarrow {\widetilde \Xbh}) \sim T_{\rm Hag} \,
 \left({\ell_s \over L}\right)^{
2(5-p)\over 7-p} \, (g_s N)^{1\over p-7} \,.}

The corresponding T-dual {\it vacuum} manifold is obtained by setting $r_0 =0$
in \smear. As before, its classical free energy is vacuum-dominated and
it is thermodynamically subleading for $p<5$, and dominating for $p\geq 5$.

  In fact,
at sufficiently low temperatures, 
 the system of `smeared' black $D0$-branes, with metric \smear,  
is unstable towards localization in the spacetime torus ${\bf T}^p$,
thus breaking the translational isometries. This was described
in \refs\rgl, and related in \refs\rcorrp\ 
 to the `fractionalization
effect', or $D$-brane wrapping, first pointed out in \refs\rfrac.   
The metric is the `array solution' of periodically identified ten-dimensional
$D0$-branes on a  ${\bf Z}^p$ grid in the $\tilde y$ directions
${\vec{\tilde y}} = {\vec n} \,{\widetilde L}$:    
\eqn\locd{
ds^2 (X_{\rm D0}) ={h_0 \over \sqrt{H_0}} \,d\tau^2 + 
\sqrt{H_0} \left( {d\rho^2 \over h_0} + \rho^2 \,d\Omega_8^2 \right),} 
with $\rho$ a ten-dimensional radial coordinate. Taking the near-horizon
limit, we  have  
 $H_0 \sim \sum_{\vec n} b_0^7    /|{\vec \rho} - {\vec n} {\widetilde L}|^7$,
$h_0 \sim 1-\sum_{\vec n} \rho_0^7 /|{\vec\rho}-{\vec n}{\widetilde L}|^7$,  
 and we can actually neglect the effect of the images. The charge radius is
 $b_0 \equiv \ell_s \,({\tilde g}_s N)^{1/7}
 $, with the T-dual asymptotic string coupling  
given by  ${\tilde g}_s = g_s (\ell_s /L)^p$, so that the charge radii
of both manifolds satisfy the matching $b_0^7 \sim {\widetilde L}^p b^{7-p}$.

The thermodynamical balance between \smear\ and \locd\ follows slightly
different patterns in the microcanonical and the canonical ensembles. 
In the microcanonical ensemble (see for example \refs\rcorrp,)
 we compare the entropies, at a fixed
value of the mass. Matching the ADM mass above extremality for both
\smear\ and \locd, we have 
\eqn\matm{
E({\widetilde \Xbh}) \sim {{\widetilde L}^p \over {\widetilde G}} \,
r_0^{7-p} \sim E(X_{\rm D0}) \sim {1\over {\widetilde G}} \,\rho_0^7,}
where ${\widetilde G}$ is the ten-dimensional T-dual Newton constant (notice
that the combination ${\widetilde L}^p /{\widetilde G}$ is T-duality invariant.)
From \matm\ we obtain the matching $\rho_0^7 \sim {\widetilde L}^p \,r_0^{7-p}$.
comparing now the entropies one finds that ${\widetilde \Xbh}$ dominates
for $r_0 \gg {\widetilde L}$, while $X_{\rm D0}$ does so for
$r_0 \ll {\widetilde L}$. 

On the other hand, in the canonical ensemble, it is the temperature of
both manifolds, \eqbexa, that has to be matched. One then finds a relation
between the horizons of the form 
\eqn\newrel{\rho_0^5 \sim {\widetilde L}^p \,\,r_0^{5-p}.} 
Now the ratio of the free energies scales  
\eqn\rfen{
{I({\widetilde \Xbh})\over I(X_{\rm D0})} \sim (5-p)\,{\beta \,{\widetilde L}^p
\, r_0^{7-p} \over \beta \, \rho_0^7} \sim (5-p) \left({r_0 \over {\widetilde L}
}\right)^{2p\over 5}  \,.}
Therefore, as long as $p<5$, the critical line 
 for the thermodynamical balance between
${\widetilde \Xbh}$ and $X_{\rm D0}$ is again
  $r_0 \sim {\widetilde L} \sim \ell_s^2 /L$. 
In terms of the gauge theory termperature, the transition locus defines  the curve
\eqn\fscover{
T({\widetilde \Xbh}\leftrightarrow X_{\rm D0}) \sim T_{\rm Hag} \,\left(
{\ell_s \over L}\right)^{5-p \over 2} \,(g_s N)^{-{1\over 2}} \,,} with
$X_{\rm D0}$, the localized solution, dominating the lower temperature,       
weaker coupling, regime. Both lines intersect a a value of the coupling
$g_s N \sim (\ell_s /L)^{3-p}$ and $T\sim 1/L$. This is very satisfying,  
since this is the temperature for the onset of finite-size effects in the
weak coupling description of the gauge theory.                      

The  transition between ${\widetilde \Xbh}$ and $X_{\rm D0}$ can be established
as a genuine large $N$, first-order  phase transition of the gauge theory, similar
to the Gross--Witten
 phase transition \refs\rgwi\ (see also \refs\rsgwi, \refs\rpolh.) 
It is the analogue, for toroidal geometry, of the `flop' described in
\refs\rwituno, \refs\rwitdos\ for CFTs in spheres.

 Again, for $p\geq 5$, the classical free energy
of ${\widetilde \Xbh}$ becomes positive because of the factor of $5-p$ in
front, as in \rfen. As a result, the solution of smeared $D0$-branes is
suppressed in the canonical ensemble, and we only have $X_{\rm D0}$ dominance
below the line \fscover.

A non-perturbative threshold develops when the T-dual dilaton in the
metric \smear, or the $D0$-brane dilaton in \locd,  becomes
large at still lower values of the radial coordinate. At this point
we must turn to an eleven-dimensional description, along the
lines of \refs\rtelaviv, where new `localization' phase transitions
can occur, as in \refs\rbfks, \refs\rhmar.
 In this paper, we are neglecting such non-perturbative
thresholds as $N$ is taken hierarchically larger than any other dimensionless
quantity in the problem.

The relation of the critical radius $r_L$ to finite-size effects can be
established heuristically through the calculation of spatial Wilson loops,
along the lines of \refs\rloop.
 Consider a Wilson loop defined at
the upper boundary of the throat with length $\ell$. In the semi-classical
approximation we must evaluate the area of the minimal surface in $X$
with boundary the Wilson loop of length $\ell$. An area-law contribution
arises whenever the minimal world-sheet `saturates' at some radius,
signaling the end of the manifold $X$. This happens at the horizon for
$\Xbh$, \refs\rwitdos, \refs\rloopt, where $r_{\rm sat} \sim r_0$.  In  general,
the     area-law contribution to
the
 Wilson loop takes the   form
\eqn\wloop{
\bra W_{\ell}  \ket_{X} \sim e^{-A_{\rm sat} /\alpha'} ,}
where the saturated area is given by the red-shifted area of the
loop in the ${\vec y}$-space:
\eqn\sata{
{A_{\rm sat} \over \alpha'}  \sim {1\over \alpha'} \,
\left({r_{\rm sat} \over b_r (r_{\rm sat}) } \right)^2 \,
\ell^2  = {(g_s N)^{2\over 7-p} \over b^2} \,
\left({ r_{\rm sat} \over b}\right)^{7-p \over 2} \, \ell^2 .}
The square root of the string tension, or effective mass gap, is
related to the saturation radius by
\eqn\rmgap{M_{\rm gap} \sim {(g_s N)^{1\over 7-p} \over b} \cdot
\left( {r_{\rm sat} \over b} \right)^{7-p \over 4} \,.}

 Since the  spatial sections of the 
manifold $\Xvac$ (or
$\Xbh$ for $r_0 \ll r_L$,) are `pinched' to stringy size  at $r\sim r_L$,
it seems natural to assume that, for the purposes of the semiclassical
approximation to the calculation of Wilson loops, the neck at $r_L$ represents
a cut-off where the minimal area world-sheet can saturate. In other words,
we can assume that the manifold $X$ `ends' at $r\sim r_L$, 
so that we may take $r_{\rm sat} \sim r_L$, leading to an induced gap     
{\eqn\gapn{ M_{\rm gap} \sim {1\over L} \,.} 
so that the saturated Wilson loop scales 
\eqn\wlt{
\bra W_{\ell} \ket \sim e^{-C \ell^2 /L^2}, }
that is, it has the interpretation of the Wilson loop hitting the
walls of the spatial box.
                
\subsec{Thermal Winding Modes and Hagedorn Effects}

We can generalize the discussion of finite-size effects to the temporal
winding modes. In this case we have Hagedorn behaviour at $r\sim r_{\beta}$
and we cannot exhibit a geometric picture of the inner region at
$r\ll r_{\beta}$, due to the lack of control over duality transformations
in the absence of supersymmetry. However, since the world-sheet physics 
is similar (infrared divergences due to light winding modes), it seems
natural to postulate a similar geometric picture in terms of a `pinching
effect' at $r\sim r_{\beta}$. This leads to a cut-off version of
$\Xvac$ at the Hagedorn radius $r_{\beta}$ which we call $\Xhag$.    
In fact, assuming that spatial Wilson loops saturate at $r_{\rm sat}
\sim r_{\beta}$,
we get an associated gap from formula \rmgap:
\eqn\rmgapt{M_{\rm mag} \sim {1\over \beta}.} 
This has the interpretation of the `magnetic mass gap' of the effective
$p$-dimensional theory arising in the high temperature limit. In this
case, we see that the gap induced by $\Xhag$ is no other than
the standard scale of Matsubara modes. This gap is to be contrasted
to the black-hole induced mass gap, with saturation at the horizon
$r_{\rm sat} \sim r_0$, which scales like \refs\rgo,         
\refs\rtelavd:  
\eqn\bhmg{
M(\Xbh)_{\rm mag} \sim {(g_s N)^{1\over 7-p} \over b} \,
\left( {b\over \beta} \right)^{7-p \over 2( 5-p)} =
\left[(g^2_{\rm YM} N) T^{7-p} \right]^{1\over 2(5-p)} .}

In fact, in the supergravity regime $g_s N\ll 1$, we tend to have
$M(\Xbh) \gg T$.  
However, before concluding that $X_{\rm Hag}$ dominates over
$\Xbh$ in the contribution to Wilson loop expectation values, we
have to determine the $\CO(e^{-N^2})$ prefactors in the expectation value
\wloop, i.e. the
relative free energies of $\Xbh$ and $\Xhag$. 

In order to estimate the free energy of $\Xhag$, 
 we shall adopt a `phenomenological'  
attitude, and assume some standard qualitative facts about the Hagedorn
behaviour of thermal strings. 
 In a world-sheet picture the Hagedorn
transition can be described \refs\rvort\  
 as a BKT  phase transition on the
world-sheet  \refs\rbkt.  
 The  world-sheet vortices  are
the vertex operators for winding modes and  their condensation means
 that   some of the winding modes become tachyonic. Due to this
 world-sheet defects, a genus-zero free energy is generated
 according to \refs\ratwitt. Translated
to the gauge theory language, this means that a planar term is induced
for the free energy of $\Xvac$ or, in other words, a classical entropy\foot{
Other discussions appear in \refs\rmave.}.

Thus, we assume that, for thermodynamical purposes,
 strong-coupling effects associated with massless  
winding modes effectively modify the $\Xvac$ manifold, capping it at
distances of order $r_{\beta}$, and generating a `stretched horizon'.\foot{
Locating the stretched horizon at the point where the local temperature
is the Hagedorn temperature is natural also in other contexts, see
\refs\rsen.}
 The
free energy of the resulting capped manifold $\Xhag$ will be calculated  
as    
\eqn\capx{I(\Xhag) = \beta M(\Xhag) - S(\Xhag).}

We approximate the entropy by the area of the stretched horizon,  
\eqn\enth{S(\Xhag) \sim {A_{\rm Hag} \over 4G_d} = {\Vol L^p \over 4G}\,
r_{\beta}^{8-p} \, \sqrt{H(r_{\beta})} \sim {\Vol  
 L^p \over 4G} \,b^{7-p \over 2} \,
r_{\beta}^{9-p \over 2} \,,}
where we have used $r_{\beta} \ll b$.
The energy of the stretched horizon can be estimated, on the basis of
dimensional analysis, and translational and rotational symmetry, to take
the form of \massex, with $r_0$ replaced by $r_{\beta}$:
\eqn\mash{
M(\Xhag) - M_{EDp} \sim {\Vol L^p \over 16\pi G}\, r_{\beta}^{7-p} \,,} 
up to constants of $\CO(1)$.   Solving for $r_{\beta}$ as a function
of the temperature, we find that the contribution of the internal energy
\mash\ to the canonical free energy \capx\ tends to make the specific  
heat negative. Let us assume for the time being that the entropy of
the stretched horizon largely dominates compared to its internal energy:  
\eqn\dome{ S(\Xhag) \gg \beta (M(\Xhag)- M_{EDp}).}
Comparing the two terms, using \rhag, this is satisfied provided $\beta \gg
\ell_s (g_s N)^{1\over p-3}$ for $p<3$, and $\beta \ll \ell_s (g_s N)^{1\over p-3}$
for $p>3$. In the conformal case, \dome\ is met when $g_s N \gg 1$. 
These assumptions will be justified later. Under these circumstances,
the resulting free energy takes the form  
\eqn\xxf{
I(\Xhag) = -N^2 \, C' \, (g_s N)^{2(5-p)(p-6) \over (7-p)^2} \, L^p \,
b^{\gamma' -p} \, \beta^{-\gamma'} + \beta \,E_{\rm vac} \, ,} 
with $C' \sim \CO(1)$ and positive, and  
$$\gamma' = {2(9-p)\over 7-p}  
.$$
 This critical exponent is harder than $\gamma_{\rm free}
=p$, and also harder than  \gammpl\ for $p<3$. It equals the conformal one for
$p=3$, and it is softer than either \gammpl\ or $\gamma_{\rm free}$ for $3<p < 6$. We
have $\gamma' = \gamma_{\rm free}$ at $p=6$. With these values of $\gamma'  $, 
given the positivity of $C'$, the manifold $\Xhag$ has positive specific
heat up to $p=7$, where the analysis breaks down.      
           
A very interesting feature of \xxf\ is that, just as in \gtfe, the dependence on
the closed string scale $\sqrt{\alpha'}$, and the effective D-brane
coupling $g_s N$, combine in such a way that the physical free energy
only depends on the Yang--Mills coupling, 
\eqn\otrav{
I(\Xhag) = \beta E_{\rm vac} - N^2 \, B' \, (g^2_{\rm YM} N)^{p-6 \over 7-p} \,
L^p \, \beta^{-\gamma'}\,.}  

\subsec{Hagedorn Censorship}

In order to determine whether the vacuum manifolds capped at the Hagedorn stretched
horizon, $\Xhag$, can dominate over the smooth $\Xbh$ manifolds in 
the canonical ensemble,    
 we compare the actions evaluated above. 
We find for the difference, 
\eqn\totf{\beta F(\beta)_{\rm planar} = I(\Xbh)_{c\ell} - I(\Xhag)_{c\ell} = 
-{\Vol L^p \over 4G} \, b^{7-p \over 2} \left( {5-p \over 2(7-p)} \, 
r_0^{9-p \over 2} - C_h \, r_{\beta}^{9-p \over 2} \right), }     
with $C_h$ a positive 
constant of order one, which summarizes our ignorance in the
treatment of the stretched horizon at $r_{\beta}$. Thus, we see that the criterion
for $\Xbh$ dominance in the large $N$ limit is $r_0 \gg r_{\beta}$, provided
$p<5$. For $p\geq 5$, the difference in 
 \totf\ is positive and $\Xhag$ dominates the
canonical ensemble to leading order. At lower temperatures, we must compare
the T-dual manifolds ${\widetilde \Xbh}$ and $X_{\rm D0}$, with the corresponding
vacuum manifolds capped at the Hagedorn radius. The Hagedorn radius and
free energy of ${\widetilde \Xhag}$ are the same as $\Xhag$ in \totf, since
T-duality of the metric in the spatial torus ${\bf T}^p$ has no effect on
the thermal direction. On the other hand, for the array solution, the Hagedorn
radius is $r_{\beta 0} \sim b_0 (\ell_s T)^{4/7}$, and the free energy scales
like $I(X_{\rm Hag0}) \sim b_0^{7/2} \rho_0^{9/2} /{\widetilde G}$.     

\ifigc\legend{Phase diagram for $p<3$ }
{\epsfxsize3.50in\epsfbox{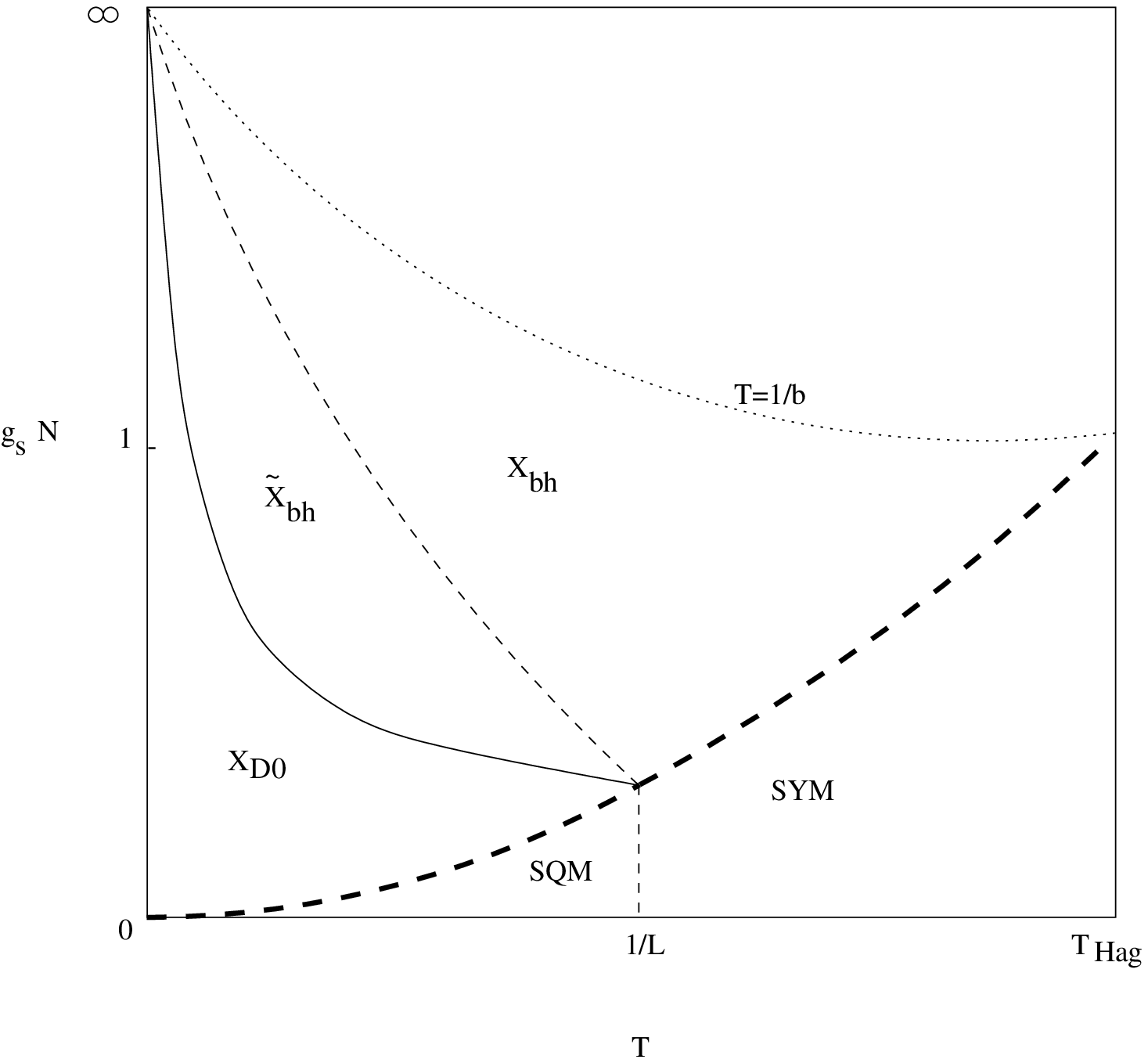}}

We now study the pattern of large $N$ `phase transitions' between 
the different descriptions of the system. In those cases where large
$\alpha'$ corrections are involved, either locally (transitions to $\Xhag$ or  
the $\Xbh\leftrightarrow {\widetilde \Xbh}$ T-duality,) or globally (transitions
to the SYM regime at $r\sim r_c$,) we cannot decide, on the
basis of our low-energy supergravity analysis,
 whether we have real large $N$ phase transitions, as
advocated recently in \refs\rli,  
or only smooth cross-over regions.  
 It was argued in \refs\rtelavd\ that the transition
between the $\Xbh$ manifold and the SYM description involves no $\CO(N^2)$
jump in the entropy, unlike the real large $N$ phase
 transition between $\Xbh$ and $\Xvac$
described in \refs\rwituno, \refs\rwitdos. However, the `flop' between
${\widetilde \Xbh}$ and $X_{\rm D0}$, described in subsection 5.1,    
is a large $N$ first-order phase transition with exact matching of
the entropies.  

The supergravity regime is defined by the requirement that the interesting
features of $\Xbh$ or $\Xhag$ be clearly separated both from the onset of
finite-size effects, and from large curvature corrections in the bulk. 
 For $p\leq 3$, this leads   to the conditions
$r_0, r_{\beta} \ll r_c$,  with $r_c =\infty$ for $p=3$. For
$p>3$, we must require $r_0 , r_{\beta} \gg  r_c$. In addition, we  exclude
from this regimes the regions of parameters where $r_0 , r_{\beta} 
\sim r_L$.

\ifigc\legend{Phase diagram for $p=3$ }
{\epsfxsize3.50in\epsfbox{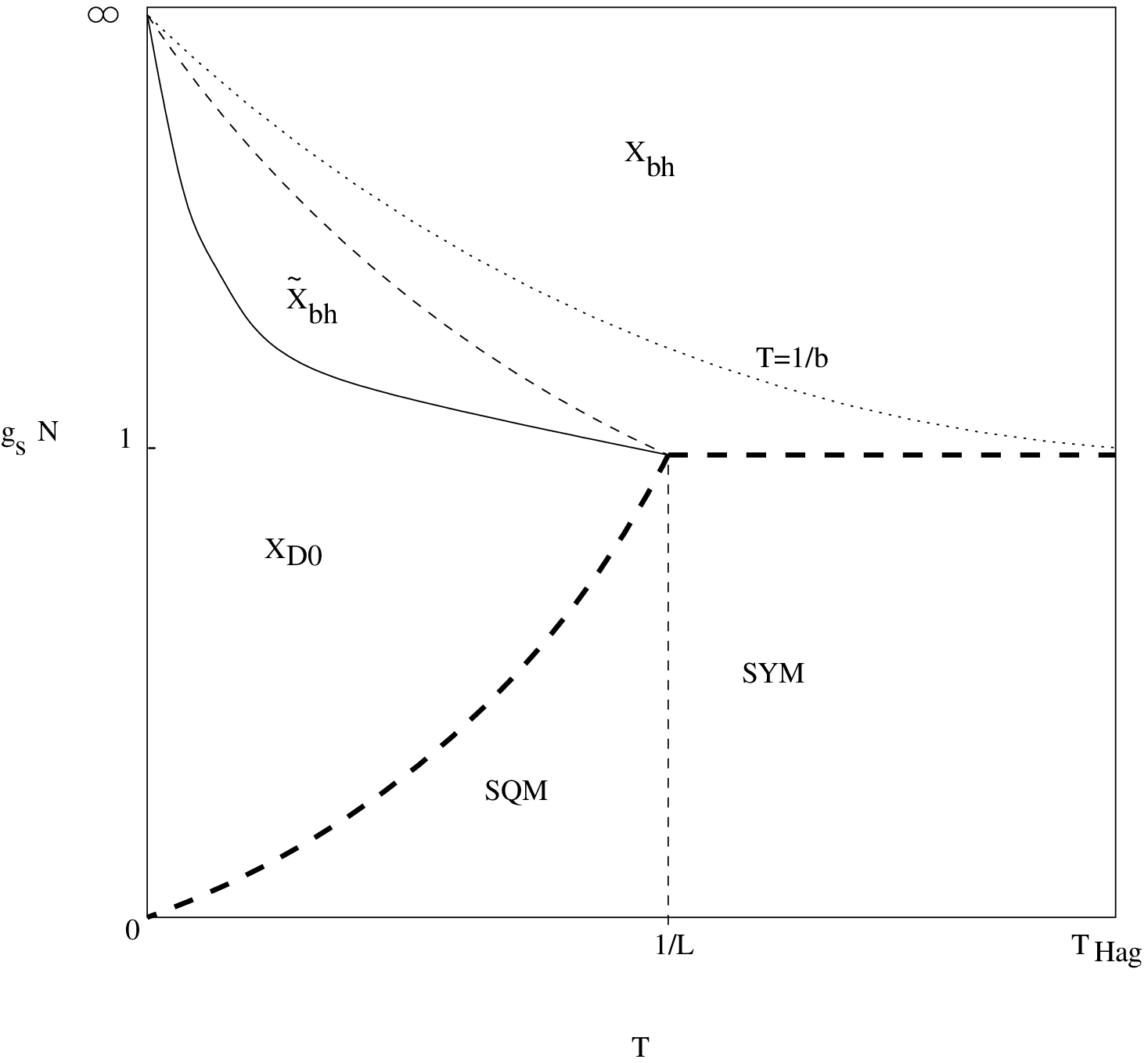}}

One can find the phase diagram by locating the cross-over regimes between
different descriptions, when they indeed dominate.
 The cross-over between the SYM and the black
hole $X_{\rm bh}$ descriptions is estimated by the condition 
$r_0 \sim r_c$, leading to the following curve in the $(T, g_s N)$ plane: 
\eqn\covers{
T({\rm SYM}\leftrightarrow X_{\rm bh})
 \sim T_{\rm Hag} \, (g_s N)^{1\over 3-p} .}
In fact, the {\it same} curve governs the transition between other, perhaps 
 subleading, descriptions,      i.e. as a function of $g_s N$ we have 
\eqn\egual{ 
T({\rm SYM}\leftrightarrow X_{\rm bh}) \sim T({\rm SYM} 
\leftrightarrow X_{\rm Hag}) \sim T(X_{\rm bh}\leftrightarrow X_{\rm Hag})
.}  
This is related to the fact that  the radii
$r_c, r_{\beta}, r_0$ become equal to each other at the same temperature. 

The transition curves due to finite-size effects are different. In the
SYM regime, we have the obvious temperature scale
\eqn\fsymm{T({\rm SYM})_{\rm f. \,size} \sim {1\over L}, }
the same that one finds for the $\Xhag$ description, determined by
$r_{\beta} \sim r_L$, i.e. 
 \eqn\fixt{
T(\Xhag\leftrightarrow {\widetilde \Xhag}) \sim {1\over L}.}
 On the other hand, for  
the black-hole manifold, the transition to the T-dual description in terms
of the smeared $D0$-branes solution occurs along the curve $r_0 \sim r_L$, or
\eqn\fins{
T(X_{\rm bh}\leftrightarrow{\widetilde \Xbh}) \sim T_{\rm Hag} \,
 \left({\ell_s \over L}\right)^{
2(5-p)\over 7-p} \, (g_s N)^{1\over p-7} \,.}
Finaly, the transition due to the localization effect between ${\widetilde \Xbh}$
and $X_{\rm D0}$ lies on the curve \fscover:
\eqn\fscover{
T({\widetilde \Xbh}\leftrightarrow X_{\rm D0}) \sim T_{\rm Hag} \,\left(
{\ell_s \over L}\right)^{5-p \over 2} \,(g_s N)^{-{1\over 2}} \,.}
Since $X_{\rm D0}$ dominates the supergravity regime in the low temperature,
weak coupling regime, the transition to the SYM description at temperatures
below \fsymm\ is determined by the curvature threshold of $X_{\rm D0}$. The 
resulting cross-over  can be obtained from \covers\ by putting $p=0$, and
substituting $g_s$ by its T-dual:
\eqn\coversl{
T({\rm SYM}\leftrightarrow X_{\rm D0}) \sim T_{\rm Hag} \, ({\tilde g}_s N)^{1\over
3} \sim T_{\rm Hag}\,\left({\ell_s \over L}\right)^{p\over 3-p} \, (g_s N)^{1\over 
3}.}
The localized $D0$-brane metric actually matches to the quantum mechanics of
the zero modes of SYM on ${\bf T}^p$, at energies below the gap $1/L$, which we
denote SQM in the pictures. According   
to \refs\rcorrp, the entropically subleading `wrapped' system, which behaves
extensively in $p$ dimensions down to very low temperatures, matches to the
(also subleading) `smeared' metric ${\widetilde \Xbh}$.
It is tempting to conjecture that the SYM theory on ${\bf T}^p$ with twisted
boundary conditions (no `toron' zero-modes) would have a behaviour more similar 
to the ${\bf S}^3$ case studied in \refs\rwituno, \refs\rwitdos, i.e. without
a phase dominated by localized $D0$-branes.  

With this information we can determine the `phase diagrams' as
in figures 4,5,6 and 7 below. 
In the figures, we have considered only `field theoretic temperatures' in the
brane theory, $T$,  up to
the string scale $T_{\rm Hag} \sim \ls^{-1}$, postponing to section 7 the
study of some aspects of the physics at boundary temperatures of the order of
the  Hagedorn temperature.
 We also plot, for
future convenience, the curve $T=b^{-1}$,  representing the maximal 
temperature
of the  complete black-brane manifold $BDp$. 
The large $N$ phase transitions due to   finite-size effects, \fscover,  appear as
 continuous lines  in the
diagrams, whereas the less understood transitions appear in dashed lines: thick 
lines for the cross-over  between the black-hole
and SYM descriptions,  the curves  \covers, \coversl,  and thin lines for
the T-duality  transition \fins.    
Notice that {\it all curves} meet at the same point in the $(g_s N, T)$ plane,
given by $T=1/L$ and $g_s N = (\ell_s /L)^{3-p}$.

\ifigc\legend{Phase diagram for $p=4$ }
{\epsfxsize3.50in\epsfbox{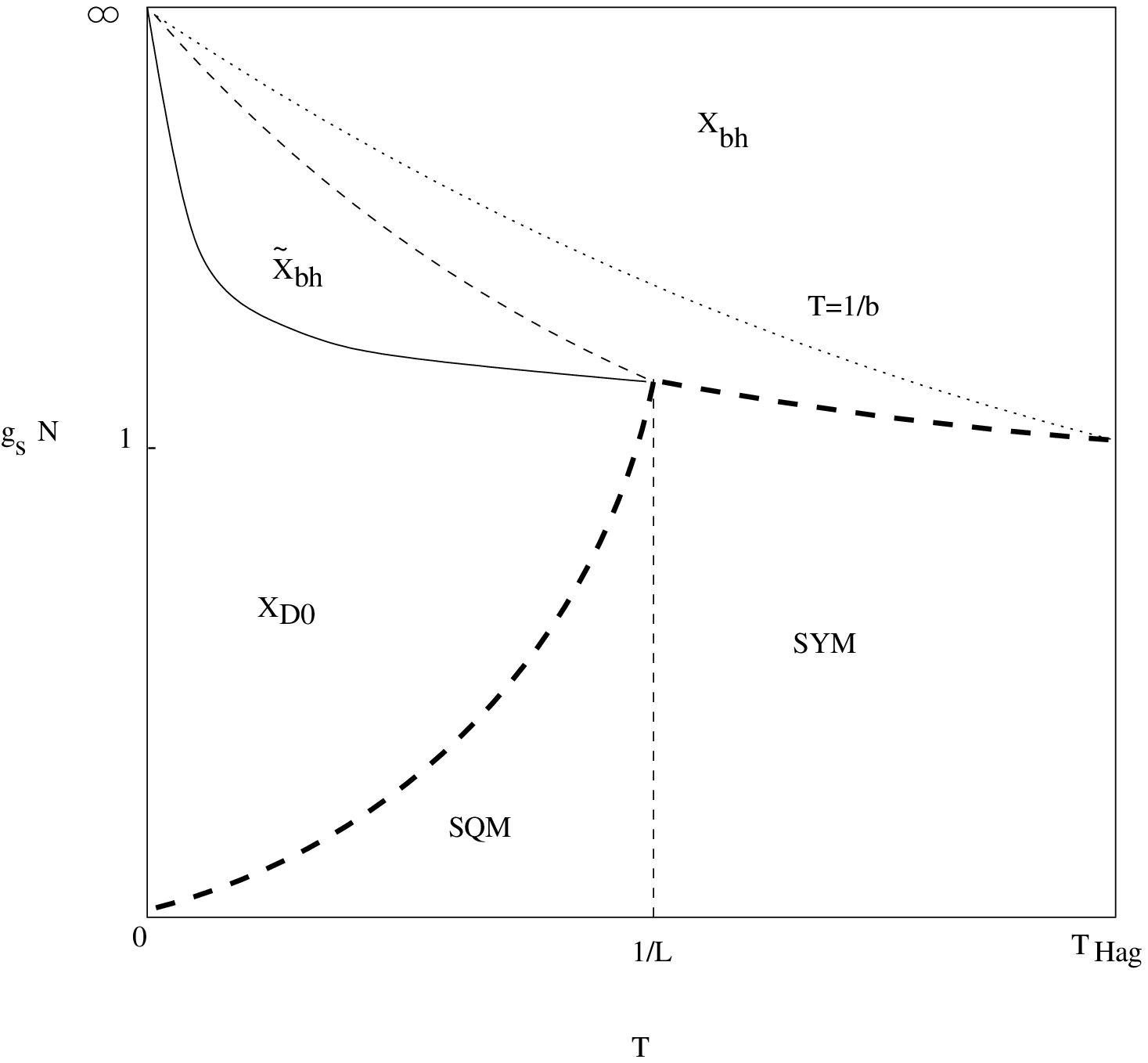}}

\ifigc\legend{Phase diagram for $p=5,6 $ }
{\epsfxsize3.50in\epsfbox{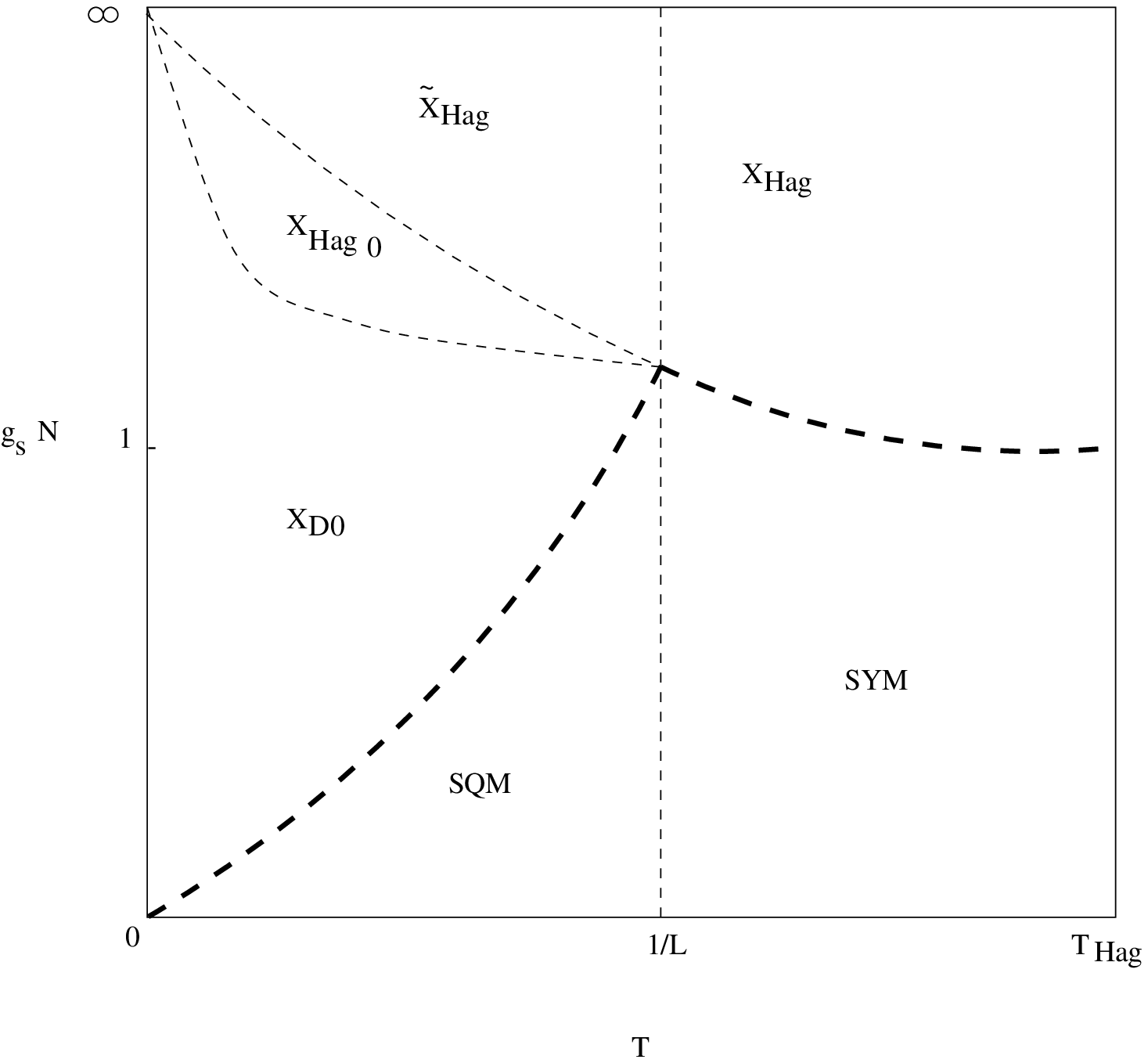}}

 The salient features of the analysis
are the following. First, the conditions of validity of \dome\ are
exactly equivalent to the condition that we are {\it not} in the
SYM regime, according to \egual. Therefore, \dome\ holds in
the region where $\Xhag$ could contribute. However, 
 one finds some hierarchies between the radii:  
$r_{\beta} \ll r_0 \ll r_c$ for the supergravity regime in $p\leq 3$,
and $r_{c}\ll r_{\beta} \ll r_0$ for the supergravity regime in 
$p=4$, with $r_{\beta} \sim r_0$ only at the cross-over to the
SYM description, i.e. $r_{\beta}\sim r_0 \sim r_c$ at the same
temperature given by \covers\ and \egual.  The result is identical for
the $X_{\rm D0}$ horizon, at $\rho=\rho_0$, versus the Hagedorn
radius on the array solution, $r_{\beta 0 }    \sim b_0 (\ell_s T)^{4/7}$.
It is also interesting to compare  the Hagedorn radii of   
the smeared and localized $D0$-brane solutions. One finds that, at very
low temperatures, $r_{\beta} \ll r_{\beta 0}$, and the cross-over
between ${\widetilde X}_{\rm Hag}$ and $X_{\rm Hag 0}$ takes place along
the curve
\eqn\hagt{
T({\widetilde X}_{\rm Hag} \leftrightarrow X_{\rm Hag 0}) \sim
T_{\rm Hag} \, \left({\ell_s \over L}\right)^{7-p \over 4} \, (g_s N)^{-{1\over
4}}.
}
This curve  also
meets all of the others at the $T\sim 1/L$ common point. 

Thus, by comparing the canonical free energies, one concludes that
 $\Xhag$ {\it never dominates} the canonical
ensemble as long as $p<5$. Thus we have a true thermodynamical 
 `Hagedorn censorship'
for these values of $p$.   

On the other hand, for $p=5,6$, the black-hole manifold $\Xbh$ has
positive free energy and is suppressed in the canonical ensemble at
high temperature. Then
$\Xhag$ dominates the supergravity regime. In particular, for $p=6$
the resulting scaling of the planar free energy is the same as in a
seven-dimensional free field theory.

In the low temperature region, there is a cross-over at $T\sim 1/L$
to ${\widetilde X}_{\rm Hag}$. At lower temperatures or couplings,
 \hagt, $X_{\rm Hag 0}$ becomes thermodynamically more favorable, and 
finally, at still lower temperatures \fscover, the smooth manifold
with positive specific heat $X_{\rm D0}$, with the array of localized
$D0$-branes, dominates and makes contact with the SQM description.     

Therefore, we do find thermodynamical domination of manifolds with Hagedorn
cut-offs for $p=5,6$. 
However, we shall find in the next section that, as soon as $p\geq 5$,
one cannot make sense of the $1/N$ corrections in terms of a gauge
theory in $p+1$ dimensions.

The Hagedorn censorship can be turned into a Hagedorn exposure if
we switch to the regime $g_s N \ll 1$. In this case $r_{\beta}$ tends
to be much larger than $r_0$. On the other hand, the whole approach
of comparing smooth manifolds breaks down, as the $\alpha'$ corrections
become of order one everywhere. This is indeed the regime where
D-brane perturbation theory is a good approximation.

If we nevertheless pursue the geometric analysis, in a formal fashion,
beyond its range of applicability,  one can see
that $\Xhag$ formally dominates ($r_{\beta} > r_0$) for sufficiently
high temperatures $\ls \ll \beta \ll b/r_c$, for
$p<3$. For $p\geq 3$ and $g_s N \ll 1$, we have $\Xhag$
dominance for all temperatures below the string scale.    
In fact, the Hagedorn cut-off manifolds overestimate the entropy with
respect to the SYM answer, i.e. if we evaluate the ratio $|I(X_{\rm Hag}) /
I({\rm SYM})|$ in the SYM regime, it is strictly larger than one for all
$p<6$, and equal to one, in order of magnitude, precisely for $p=6$.
In any case, the resulting high-temperature string tension in the
effective $p$-dimensional theory, does not agree with the perturbative
determination, and therefore  it is hard to determine the relevance of 
this geometric picture at $g_s N \ll 1 $.

\newsec{Holography Criterion at One Loop}  

As pointed out in \refs\rbr, a necessary condition for holography to
hold in an operational sense is that, by heating up the gauge
theory, we do not uncover the full ten- or eleven-dimensional
structure of the supergravity description. This can be tested
by estimating the one-loop free energy in the supergravity side, which
should be interpreted as the leading $1/N^2$ correction on the 
gauge theory side.

In a space with locally varying Kaluza--Klein thresholds, the number
of effective dimensions is given by the number of `circles' with
proper radius larger than the local inverse temperature. In a region
$X(d_{\rm eff})$ 
where this number 
of effective dimensions is fixed, $d_{\rm eff} = {\rm dim}(X(d_{
\rm eff}))$,  we can estimate the one-loop
free energy by a red-shift \refs\rhawkp, or WKB formula \refs\rthooftb:  
\eqn\rse{ 
\beta F_{\rm WKB} \sim -\int_{X(d_{\rm eff})} d{\rm Vol} \,\left({1\over 
\beta \sqrt{g_{00}}}  
\right)^{d_{\rm eff}}     
+ \beta E_{\rm vac}\,.}
In other words, the one-loop free energy in curved space is approximately
extensive (in the standard sense of linearity) in the so-called optical
volume \refs\ropvol, i.e. the volume measured in the optical
metric\foot{Notice
that the optical metric ${\widetilde g}_{\mu\nu} \equiv g_{\mu\nu}/g_{00}$
is invariant under conformal rescalings of $g_{\mu\nu}$. This means that
it is the same for the string or Einstein frames. In fact, it is the
Einstein-frame metric the relevant one for one-loop computations in
supergravity, because the dilaton has a negative kinetic energy in
the string frame metric.},
\eqn\roptmet{ {\widetilde g}_{\mu\nu} = {g_{\mu\nu} \over g_{00}}\,.} 
For a manifold with several thresholds with different effective dimensions
$d_i$, we have
\eqn\feap{ \beta F(\beta)_{\rm 1-loop} \sim -\sum_{X_{d_i}} A_i \, 
\beta^{-d_i} \, {\widetilde {\rm Vol}} (X_{d_i}) +\beta E_{\rm vac}\,,}
where the constants $A_i$ are related to the number of degrees of freedom. In
the present case, $A_i \sim \CO (1)$ in the large $N$ limit. 
Notice that the real temperature scaling of the contribution from a given
region $X_{d_i}$ could differ from $T^{d_i}$, because of the temperature
dependence of the optical volume, hidden in the definition of the limits
of the region $X_{d_i}$ itself.  
  
The expression \feap\ was analyzed in \refs\rbr\ for the conformal cases
of the form $\Xvac = AdS_{d+1} \times M_{D-d-1}$, with $M_{D-d-1}$ a compact 
manifold of constant curvature and radius $b$. Examples include the CFT
on the $D3$-branes world-volume for $D=10$, and the $M2$- and $M5$-branes
world-volume CFTs for $D=11$. In all these cases, the black-hole manifold
$\Xbh$ takes the form of an AdS Schwarzschild black-hole extended over
$M_{D-d-1}$ and with horizon radius $r_0 \sim b^2 T_0 $, with $T_0$ the local
temperature at the centre of the $\Xvac$ space. The condition of decoupling
of the Kaluza--Klein threshold of the compact manifold $M_{D-d-1}$ leads to an  
 asymptotic region with $d+1$ effective dimensions, and a `core' region
with $D$ effective dimensions. The cross-over is at $r_{\rm dec} \sim r_0$, so
that $\Xbh$ is completely $d+1$ dimensional for thermal purposes, while
$\Xvac$ has a `core' for $r< b^2 T_0 $ where it is effectively $D$-dimensional.  
Computing now the optical volume of the different regions and plugging it
in the general expression \feap, one finds that $\Xvac$ is non-holographic,
with a $D$-dimensional scaling behaviour $\sim T^{D-1}$, coming from the
core. The asymptotic region of both $\Xvac$ and $\Xbh$ yields {\it always}
a holographic contribution to the free energy of the form $\sim T^{d-1}$. 
Hence, one concludes that $\Xbh$ dominance guarantees holographic behaviour
as characterized by the power dependence of the $1/N^2$ corrections to the
free energy.   

We generalize now the discussion to the the more 
general non-conformal cases treated in this paper. In fact, 
 we can carry out the analysis for all the 
 branes of low-energy superstring theory
at once, using the general solutions of \refs\rdp,  
\eqn\mgen{ds^2 = {1\over H^{\delta(a+1)}} \left(h \,d\tau^2 + d{\vec y}^{\ 2} 
\right) + 
{1\over H^{\delta (a-1)}} \left({dr^2 \over h} + r^2 d\Omega_{8-p}^2 \right), }  
where $a=0$ for $Dp$-branes,  $a=-1$ for NS fivebranes and $a=1$ for
the fundamental string solution. The constant $\delta$ is       
$$
\delta = {1\over 2a^2 - (3-p) a +2}
\,.$$
  Notice that $\delta = 1/2$ for all $Dp$-branes as well as the NS five and
one-branes, i.e. for all BPS branes of the ten-dimensional 
 type II supergravity or truncations
thereof.

The decoupling of the angular sphere ${\bf S}^{8-p}$, with a position
dependent radius given by
\eqn\rlocal{
R_{\rm local} = {r \over (\sqrt{H})^{\delta (a-1)}}, } 
 depends on the ratio between this radius and the
local inverse temperature
$$
\beta_{\rm local} = {\beta\sqrt{h} \over (\sqrt{H})^{\delta (a+1)}}.
$$
The ratio scales in the asymptotic, large $r$ region as
\eqn\rat{
{\beta \sqrt{h} \over r} \, H^{-\delta} \sim {\beta \over b^{(7-p)\delta} \,
r^{1-(7-p)\delta} }.} 
Therefore, if $1-\delta (7-p) = -(5-p)/2 <0$, the angular sphere ${\bf S}^{8-p}$
decouples up the throat. In this case $(p<5)$, the effective holographic
manifold for thermal purposes is $X_{p+2}$, obtained from the full ten-dimensional
manifold $X$, by dropping the angles, i.e. we remain with the $(\tau, {\vec y}, r)$
components of the metric, or $AdS_{p+2}$ with a variable radius $R_{\rm local}$. 

If $1-\delta (7-p) =0$, or $p=5$, the decoupling depends on the ratio
$\beta/b$. 
For $\beta \gg b$,  the angular
sphere decouples, but it must be kept in the opposite, high temperature
regime $\beta \ll b$.      

On the other hand, if $1-\delta (7-p) >0$, or $p>5$, then the angular
radius is larger than the local inverse temperature up the throat, and
these extra dimensions {\it do not decouple} from the thermal partition
function. 

In all cases, except the marginal one at $p=5$, 
 the radial cut-off defining the asymptotic region can
be taken as the radius at which $\beta_{\rm local} \sim R_{\rm local}$,
or
\eqn\rct{
r_{\rm dec}^{1-\delta (7-p)} \sim {\beta \over b^{(7-p)\delta}}.
}
The full optical metric in the asymptotic region, with $h\sim 1$, is
\eqn\oppp{{\widetilde ds}^2 (X)
 \rightarrow d\tau^2 + d{\vec y}^{\ 2} + H^{2\delta}
(dr^2 + r^2 d\Omega_{8-p}^2 ), }
with optical volume between two limits given by 
 \eqn\opvl{\eqalign{ \left[{\widetilde{\rm Vol}} (X)\right]^{r_{\rm max}}_{r_{\rm
 min}} =
& \beta \, L^p \, \Vol \,\int H^{\delta (9-p)} rdr \cr  =
&\beta\, L^p \, \Vol \, b^{\delta(7-p)(9-p)}
 \, \left[
{r^{(9-p)(1-\delta (7-p))} \over (9-p)(1-\delta (7-p))} \right]^{r_{\rm max}}_{r_{\rm
min}} \,. }
}
  On the other hand, the optical metric of $X_{p+2}$ is the same as
in \oppp, after dropping
the angles (the $d\Omega_{8-p}^2$ term), and the optical volume 
\eqn\opvp{
 \left[{\widetilde {\rm Vol}} (X_{p+2})
\right]^{r_{\rm max}}_{r_{\rm min}} = \beta\, L^p \, \int_{r_{\rm min}}^{r_{\rm max}}
 H^{\delta} \,dr \sim
\beta \, L^p \, b^{\delta (7-p)} \, \left[{r^{1-\delta (7-p)} \over 1-
\delta (7-p)} \right]^{r_{\rm max}}_{r_{\rm min}}\,.}

Notice that, in both cases,
 the quantity governing the divergence properties of the optical volume at
large radius  is
$$
1-\delta (7-p) = - {5-p \over 2}
.$$
 
It is  interesting to notice that the
`holography index' $1-\delta (7-p)$ is also governing the
leading, $\CO (N^2)$
thermodynamical properties of the black-brane manifold $\Xbh$.
By applying \eqbeta\ to the general
metric \mgen, one finds exactly the same result for the relation between
temperature and horizon location. In the $r_0 \ll b$ regime:
\eqn\bbt{\beta =
    {4\pi \over 7-p} \,r_0 \,\left[H(r_0)\right]^{\delta}
\sim b^{\delta (7-p)} \, r_0^{1-\delta (7-p)} \,. }
Thus, the correlation between the finiteness of the optical volume and
the positive specific heat of the classical black-brane solution is a
property of all black branes, independently of their RR or NS character.
Incidentally, \bbt\ shows that $r_0 \sim r_{\rm dec}$, with $r_{\rm dec}$ the cut-off
defined in \rct\ signaling the decoupling  of the
angular sphere. This is a property of the exactly conformal throats
\refs\rbr, which generalizes to  other $Dp$-branes.

 For $p< 5$,
the $\CO(N^2)$ thermodynamics is dominated by $\Xbh$ at sufficiently high
temperatures. We can then set $r_{\rm min} = r_{\rm dec}\sim r_0$, and the resulting 
 one-loop partition function picks a contribution from large radius
of the form 
\eqn\conv{
\beta F(p<5)_{\rm 1-loop}
 \sim -T^{p+2} \, \left[{\widetilde {\rm Vol}} (X_{p+2})\right]_{r_0}^{\infty} 
 \sim - ( L\, T)^p \,.}    
Here we have set $r_{\rm max} = \infty$, since the integral converges there. It is
remarkable that we find a fully holographic answer, characteristic of a
$p+1$ dimensional free gas.

For the borderline case $p=5$, the planar approximation to the canonical ensemble
leads to $\Xhag$ dominance, hence we must set $r_{\rm min} \sim r_{\beta}$.
 There is a critical temperature, $T_{\rm dec} \sim 1/b$,
 for the global decoupling
throughout the throat of the 
angular sphere ${\bf S}^3$, and we obtain   
\eqn\fcincl{
\beta F(p=5)_{\rm 1-loop} \sim -T^{5+k} \, \left[{\widetilde {\rm Vol}}
 (X_{5+k})\right]_{r_{\beta}}^{r_{\rm max}} 
\sim - (L\, T)^5 \, \left({T\over T_{\rm dec}}\right)^{k-1} \, {\rm log}\,
\left( {g_s N \,T_{\rm dec}^3 \,r_{\rm max} \over T^2 }\right) ,}  
with $k=2$ for $T\ll T_{\rm dec}$, and $k=5$ for $T\gg T_{\rm  dec}$. In both cases,
we do get a violation of holography from the dependence on the gauge
theory ultraviolet cut-off $r_{\rm max}$. At moderately high temperatures $(L^{-1} \ll 
T\ll T_{\rm dec})$,  
we find a seven-dimensional gas scaling, while the full ten-dimensional scaling
is uncovered for temperatures above the angular decoupling threshold.

Finally, for $p=6$, we must use the full optical volume of the
$X_{10}$ manifold, as the angular sphere does not decouple at sufficiently large
radius.
 We have argued in the previous section that $\Xhag$ dominates the canonical 
ensemble 
at high temperature $T\gg 1/L$.   
We can also define a `decoupling temperature' depending on whether the
Hagedorn radius is larger or smaller than the angular decoupling radius. This
temperature is thus defined by the condition $r_{\beta} \sim r_{\rm dec}$
and is given by
\eqn\decs{
T_{\rm dec} \sim {(g_s N)^{2/3} \over b} \sim {T_{\rm Hag} \over (g_s N)^{1/3}}.} 
We have two regimes. If $T\ll T_{\rm dec}$ the WKB estimate is
\eqn\seisl{
\beta F(p=6)_{\rm 1-loop} \sim -T^8\,\left[{\widetilde {\rm Vol}}
 (X_{8})\right]_{r_{\beta}}^{r_{\rm dec}} - T^{10}\, 
\left[{\widetilde {\rm Vol}} (X_{10})\right]_{r_{\rm dec}}^{r_{\rm max}} 
\,.}
The first term gives a holographic contribution, of the form
$$
-T^7 \, L^6 \, \sqrt{b}\,\left(\sqrt{r_{\rm dec}} -\sqrt{r_{\beta}}\right) \sim
-(L\,T)^6 \,\left( 1-(T/T_{\rm dec})^3 \right),
$$
because the term with ten-dimensional scaling $\sim T^9$ is small for
$T\ll T_{\rm dec}$, compared to the holographic one, scaling with the
law $\sim T^6$. On the other hand, the second term in \seisl\ leads to
unsuppressed ten-dimensional behaviour, again from the dependence on
the ultraviolet cut-off $r_{\rm max} $:
$$
-T^9 \, L^6 \, b^{3/2} \,\left(r_{\rm max}^{3/2} - r_{\rm dec}^{3/2} \right) \sim 
-T^9 \, L^6 \,(b\, r_{\rm max} )^{3/2} + (L\,T)^6 \,.
$$

In the high temperature regime $T\gg T_{\rm dec}$ we have
\eqn\seish{
\beta F(p=6)_{\rm 1-loop} \sim -T^{10} \,
\left[{\widetilde {\rm Vol}} (X_{10})\right]_{r_{\beta}}^{r_{\rm max}} 
\sim -T^9 \, L^6 \,(b \,r_{\rm max})^{3/2} + (L\,T)^6 \,\left({T\over T_{\rm dec}}
\right)^9 \,.}
In this case, even the term with no cut-off dependence shows pathological
behaviour, with an unsuppressed power law $\sim T^{15}$.

In summary, for $p<5$ the planar specific heat is positive, making
  possible a gauge theory interpretation of the gravitational thermodynamics.
Precisely in this range the first $1/N^2$ or one-loop correction to the
free energy scales like 
 a massless free gas in $p$ spatial dimensions.
On the other hand, for $p\geq 5$ the optical volume diverges, and we
pick a term at $r=r_{\rm max}$ with ten-dimensional temperature dependence
$\beta F \sim -T^9$, thus leading to a breakdown of holography.

 At this
point, it should be noted that the reported violation of holography
in the $p=5$ case is a rather marginal effect, since the dependence
on the ultraviolet cut-off is only logarithmic. It could be that our
criterion is too crude in this case, and a more refined criterion would
recover $p=5$ as a holographic system.    

The phenomenon described here is
 similar to the one found in conformal AdS backgrounds
in \refs\rhoo\ and \refs\rbr. However, in those cases, the violations
of holography where confined to the interior of the $X$-manifold, and
in particular the asymptotic region was always holographic. Here, 
the asymptotic region is  the one that shows ten-dimensional scaling.

The natural interpretation of the holography breakdown for $p\geq 5$
would be, as in \refs\rss, \refs\rsmal,
 as a breakdown of the decoupling between
throat physics and ten-dimensional gravity in the asymptotically flat
region. In other words, the procedure of blowing up the throat, which
is, according to \refs\rmalda, dual to the decoupling of string thresholds
in the D-brane theory, is not stable under quantum corrections in the
gravity description.

\newsec{Open-String Thresholds on the Brane} 
 In the previous sections, we have analyzed some aspects of the thermodynamics
of the SYM/SUGRA correspondence, in the strict low energy limit of
\refs\rmalda. From the point of view of the weakly coupled description
on the brane, this means that the  open-string oscillator modes are decoupled
and one only keeps those degrees of freedom corresponding to SYM theory
in $p+1$ dimensions. From the point of view of the brane geometries,
the same procedure isolates the throat geometries or $X$-manifolds. In
particular, these geometries can be continued past the `neck position'
at $r\sim b$. The U.V./I.R. correspondence \refs\rmalda, \refs\rws\
implies that the radial coordinate acquires the interpretation of a 
renormalization group scale. Therefore, the `elimination' of the 
neck by the blow-up procedure is interpreted as a tool to complete
the system in the ultraviolet using only Yang--Mills degrees of
freedom, wherever that can be done. 

It seems natural to suspect that, keeping the neck in place at radial
coordinates of order $r\sim b$, should be somehow related to the introduction
of the string scale threshold of the brane theory. A precise statement in
this direction was made in \refs\rigg, where corrections to conformality
induced by Dirac--Born--Infeld terms (DBI)
 in the open-string effective action where
related to the precise matching of wave functions at the finite position of
the neck in the full $D3$-brane metric.

We can understand this in very general terms via a
 simple modification of the scaling argument of section 4, indicating  
that the neck position at $r\sim b$ is giving us some information about
$\alpha'$-suppressed (i.e. DBI-type) processes in the
$Dp$-brane theory at large $g_s N$. If, instead of the scaled-up
geometries $\Xbh$ and $\Xvac$, we use the complete $EDp$ and
$BDp$ metrics in \dpsol\ and \bbrane, we cannot completely scale out 
the charge radius $b$ from the metric.
 The functions left undetermined by the scaling
argument, such as the one appearing in eq. \scalr, become functions of
two variables, i.e. $f(z^{2\alpha}/z_0^{2\alpha}, z^{2\alpha} /
b^{2\alpha})$ or, translating back to the radial variable $r$:
$f(r^{7-p}/r_0^{7-p}, r^{7-p}/b^{7-p})$. The result is that each
term in \cor\ is multiplied by an unknown function of the
ratio $r_0^{7-p} /b^{7-p}$. 

If we now do a power-expansion in this ratio, the $k$-th order
term scales like
\eqn\salpha{
\left({r_0 \over b}\right)^{k(7-p)} = \left(\alpha'^2 \,M_{\rm mag}^4 
\right)^k \,,}
where $M_{\rm mag}$ is the `magnetic mass gap' in eq. \bhmg. We see
that the cancellation \gtfe\ of the $\alpha'$ dependence in favour of gauge
theory quantities is not complete. Indeed, the leading correction 
scales like $M_{\rm mag}^4$, i.e. the correct one for the expectation
value of the $F^4$ combination in the DBI action.

In this section, we adapt the SUGRA thermodynamical analysis of the previous 
sections to the case where one keeps the neck in place. We also take a peek
into the Hagedorn regime of open strings on the $Dp$-brane at weak coupling,
and obtain $p=5$ as a critical dimension also from this point of view.   

\subsec{Keeping the  `Neck' in Sight}
The main thermodynamical  
effect of keeping the neck of the brane geometries in place
is clearly seen in fig. 1. Unlike $\Xbh$, the full black-brane geometry
$BDp$ has always a Schwarzschild regime with negative specific heat and
$r_0 \sim \beta$. This means that, at a temperature of order $T_b \sim b^{-1}$,  
the canonical free energy of $BDp$, with $p<5$,  becomes positive, and is therefore 
suppressed in the canonical ensemble. In the  approximation of smooth geometry, we
have then a first-order transition to a high temperature regime dominated
by the extremal $Dp$-brane geometry $EDp$.
 In this approximation, there is no $\CO(N^2)$
entropy in the high temperature phase, and one has exactly the same behaviour
as in \refs\rwituno, \refs\rwitdos, with the roles of high and low temperature
interchanged. 

However, including the effects of thermal winding modes, we know that
$EDp$ acquires a cut-off at the Hagedorn radius $r_{\beta}$, generating
a stretched horizon and thus a planar entropy. Therefore, the picture of 
the `phases', as presented in figs. 4-7, changes simply by the addition of
a $\Xhag$-dominated phase at high temperatures, above the line $T=b^{-1}$. 
We see that one can obtain `Hagedorn exposure' at sufficiently high temperatures
in the gravitational description, provided the throat neck is kept in place. 
This description can be continued up to string scale temperatures. When
$T\sim T_{\rm Hag}$, the Hagedorn radius becomes of the order of the
neck size, $r_{\beta} \sim b$, and the stretched horizon comes out of the
throat. At this point it is difficult to imagine any consistent decoupling
between the ten-dimensional bulk and the world-volume physics.    

On the other hand, for $p\geq 5$ we have $\Xhag$ dominance already in
the blow-up limit, and no particular change of the planar thermodynamics
 takes place by limiting
the end of the throat in the ultraviolet regime. 

To conclude this discussion, we
  should remind the reader that we are considering the canonical ensemble
throughout the paper. In a microcanonical description, the Hagedorn cut-off
$EDp$ metric never dominates over $BDp$, because their masses can only
be matched when $r_0 \sim r_{\beta}$, that is, at the transition line  
to the SYM description.  Indeed, the microcanonical description of the
system with the neck in place is equivalent to the one given   in \refs\rcorrp.   

\subsec{Hagedorn Regime at Weak Coupling and $p=5$ Again}

 Let us consider now the problem of the  Hagedorn transition at weak coupling,
using more traditional methods.
It was mentioned at the end of section 5 that, for  $g_s N \ll 1$, the
radius   
 $r_{\beta}$ tends to be much larger than $r_0$  and  one can think that the
BKT phase transition takes place before $\Xbh$ starts to dominate.
 Of course  the whole approach
of comparing smooth manifolds breaks down, as the $\alpha'$ corrections
become of order one everywhere. This is indeed the regime where
D-brane perturbation theory is a good approximation.  It will be
interesting to see if in this approach we can see that $p=5$ is
special again.

 In the
 limit $g_s \rightarrow 0$, with $g_s N$ fixed and small, we can decouple the
$Dp$-brane world-volume from closed strings in the ten-dimensional flat
bulk space-time, {\it at all energies}, i.e. with no assumptions on the
value of $\alpha'$. In this limit, the bare mass of the $Dp$-branes also
diverges, and we can neglect brane creation processes in the thermal gas,
i.e. we focus on the open-string thermal gas confined in the $p+1$ dimensions
of the world-volume. The one-loop free energy of open strings stuck at the
 $Dp$-brane 
takes the form
\eqn\sevenone{
\beta F =- \half \int_0^{\infty} {dt \over t} \Tr_{{\cal H}_{\rm open}}
 (-1)^F \, e^{-t \Delta_{\rm open}} }
with ${\cal H}_{\rm open}$ denoting  the open string Hilbert space
and the free inverse  propagator
\eqn\seventwo{{1\over \alpha'} \Delta_{\rm open} =  ({\vec p}_N)^2 + {4\pi^2 n^2
 \over
\beta^2} - {a\over \alpha'} + {\rm osc}
. }
Here ${\vec p}_N $ is the momentum in the $p$ spatial Neumann
directions and $n$ is the quantum of Matsubara frequency.
   The constant $a$ is the normal ordering intercept: $a=1$ for
bosonic strings, while $a_{\rm NS} = 1/2$ for Neveu--Schwarz spin
structure and $a_{\rm R} =0$ for the Ramond sector, in the case of superstrings.
The Hagedorn behaviour in this representation comes from the ultraviolet region
of the open string world-sheet $t\sim 0$, and it is governed by the
asymptotic growth of the density of states. We can isolate it by performing
a modular
transformation to the closed string channel. In this case the previous
annulus amplitude becomes a closed string cylinder representing a propagator
between boundary states \refs\rgreenbs. \eqn\seventhree{
\beta F(\beta) = - \bra B, x |\,P_{\rm GSO}\, {1\over \Delta_{\rm closed}} | B, x
\ket
.}
Now the closed string inverse propagator is
\eqn\sevenfour{
{2\over \alpha'} \Delta_{\rm closed} =  {\vec p}^{\ 2} + {4\pi^2 n^2  \over
\beta^2} + {\beta^2 \ell^2 \over 4\pi^2 \alpha'^2} - 2\cdot{ a+{\overline a}\over
 \alpha'}
 + {\rm osc}
.}
In this case,
 ${\vec p}$ explores all spatial dimensions, $n$ is as before the Matsubara
frequency, now of the closed strings, and $\ell$ is the thermal winding number of
the closed strings.  The boundary state $|B, x\ket$ contains an oscillator part
that enforces the Neumann or Dirichlet boundary conditions depending on the
coordinate. The zero-mode part is such that the momentum flow is zero for
Neumann directions, including the thermal one,
 and the position is fixed for Dirichlet directions, that is
\eqn\sevenfive{
|B, x\ket_{\rm zero\,mode} = \int {d{\vec p}_D } \,e^{i{\vec p}_D \cdot {\vec x}}
 \;
|{\vec p}_D \,;\,{\vec p}_N = n=0 \ket
.}
Notice that the closed strings are free to wind in the thermal direction
(i.e. we sum over $\ell$).
 
 Using this form of the boundary state we can write the free energy in the
closed channel as
\eqn\sevensix{
\beta F(\beta) \sim -\int d{\vec p}_D \sum_{s,\ell} {|\bra B|s,\ell\ket|^2
 \over
{\alpha' \over 2} \left({\vec p}_D^{\ 2}  + M^2_{(s,\ell)} \right) }
,}
where $s$ is an index for the oscillator states of the closed strings and
$M_{(s,\ell)}$ is the mass of closed string states in the dimensionally
reduced theory. The Hagedorn transition is here an infrared effect associated
to a new massless mode of the {\it closed} channel
 $M^2 \sim \beta-\beta_c$ at winding number $\ell =1$. The
critical inverse temperature is
$$
\beta_c^2 = 8\pi^2 \alpha'\, (a+{\overline a})
.$$
Using
a Schwinger proper time representation for the propagator, we need to find the
singularity coming from $\tau=\infty$ in
\eqn\sevenseven{
\beta F(\beta) \sim -
 \int d{\vec p}_D \sum_{s,\ell} |\bra B|s,\ell\ket|^2 \int_0^{\infty}
d\tau \,e^{-\tau {\alpha' \over 2}({\vec p}_D^{\ 2} +M^2_{(s,\ell)})}
.}
Doing the ${\vec p}_D$ integral and picking only the infrared part, we are led to
\eqn\seveneight{
\beta F(\beta)_{\rm sing} \sim \int_{1}^{\infty} {d\tau \over \tau^{9-p \over 2}}
 \,
e^{-\tau \cdot {C\over \sqrt{\alpha'}}
 (\beta-\beta_c )} \sim \left({\beta-\beta_c \over
\sqrt{\alpha'
}} \right)^{7-p \over 2}  \;\Gamma((p-7)/2 \,;\, C(\beta-\beta_c)/\sqrt{\alpha'} )
.}
Before analyzing this expression for free energy let us show that it
 simply describes the free energy of a gas of particles moving in a
$p+1$ dimensional space-time  with the
Hagedorn  spectrum
$$
 \rho(m) \sim m^{-9/2}\exp(\beta_{c} m)
.$$
 This is the spectrum of open superstrings on a
$p$-brane. Calculating the free energy
 for bosons one gets
\eqn\sevenboson{
\eqalign{-\beta F_{b} = \ln Z_{b} &=
- {L^p\over (2\pi)^{p}} \int_{m_{0}}^{\infty}
 dm\, \rho (m) \,\int d^{p} k\, {\rm log}\,\left[ 1 -
\exp\left(-\beta\sqrt{k^{2} + m^{2}}\right)\right] \cr 
&={L^p \over (2\pi)^{p}} \int_{m_{0}}^{\infty}  dm \,\rho (m)
\sum_{n=1}^{\infty} {1 \over n} \int d^{p} k
\,\exp\left(-n\beta\sqrt{k^{2} + m^{2}}\right)~~~~
,
}}
 and for fermions
\eqn\sevenfermions{
\eqalign{- \beta F_{f} = \ln Z_{f} &=
{L^p \over (2\pi)^{p}} \int_{m_{0}}^{\infty}
 dm\, \rho (m)\, \int d^{p}k \,{\rm log}\,\left[ 1 +
\exp\left(-\beta\sqrt{k^{2} +  m^{2}}\right)\right] \cr  
&={L^p \over (2\pi)^{p}} \int_{m_{0}}^{\infty}  dm \,\rho (m)
\,\sum_{n=1}^{\infty} {(-1)^{n+1}\over n} \int d^{p}k
\,\exp\left(-n\beta\sqrt{k^{2} +  m^{2}}\right)~~~~
,
}}
where $m_{0}$ is the infrared cut-off, which is usually of  the same
order
 of magnitude as  $\beta_{c}$.
Because we are looking at  $\beta \sim \beta_{c}$,   practically
  all particles  can be considered as non-relativistic ones and
 we can   rewrite
$\exp\left(-n\beta\sqrt{k^{2} + m^{2}}\right)$
as $\exp(-n\beta m-n\beta k^{2}/2m)
\left(1+O(n\beta k^{4}/m^{3})\right)$;
it is easy to check  that the  neglected
 terms are of   order   $(n\beta m)^{-1}\ll 1$.
Integrating over $k$  we finally get
\eqn\freeenergy{
- \beta F_{b(f)}
={L^p \over (2\pi)^{p/2}\beta^{d/2}} \int_{m_{0}}^{\infty}  dm
\, m^{p/2} \,\rho (m)\sum_{n =1}^{\infty}
{(-1)^{n+1} \over n^{p/2+1}} \exp(-n\beta m)
.}
 
It is clear that in the vicinity of the Hagedorn temperature
 $\beta \rightarrow \beta_{c}$
  one can  neglect all terms in the sum with $n \geq 2$ (they will
 give singularities at $T = n T_{c}$)
 and  there is no difference between bosons and fermions in the
 leading singular term with $n=1$. After integrating over $m$ we
get the  same answer
\eqn\seveneightagain{
\beta F(\beta)_{\rm sing} \sim   \left({\beta-\beta_c \over\sqrt{\alpha'
}} \right)^{7-p \over 2}  \;\Gamma((p-7)/2 \,;\, C(\beta-\beta_c)/\sqrt{\alpha'} )
,}
 where the  constant $C$ is related to the  mass parameter $m_0$.
 Let us note that these two approaches are dual because in the first
case we integrated over momenta in the  Dirichlet 
directions and in the second
case over the momenta in the Neumann directions (along the brane).
For further discussion of such thermal duality properties see
\refs\rma.  
Now one can see that for  $p/2  > 7/2$ the
  free energy  has a power singularity   when  $\beta \rightarrow
\beta_{c}$,
 whereas  in the case $ 7/2 > p/2$ it is  finite at
 $\beta = \beta_{c}$.  For $p = 7$ one has a logarithmic
 singularity $ \beta F \sim \ln\left((\beta-\beta_{c})m_{0}\right)$.
 That is, for $p\geq 7$ the critical behaviour is similar
to the case of normal open strings, in that {\it all} thermodynamical functions
diverge at the critical point. For $p<7$ the free energy is finite at the
transition. The internal energy diverges for $p\geq 5$ and the specific heat
diverges for $p\geq 3$.
 
This means that the Hagedorn temperature starts being `maximal' for $p=5$,
as one has to supply an infinite amount of energy to raise the temperature
to that level. However, when the free energy stops diverging, the canonical
and microcanonical ensembles could lose equivalence. This is related to
the possibility that the specific heat becomes negative. 
The classic analysis of ref.  
 \refs\rfc\ can be generalized  
 to our case,  and one obtains, in terms  
 of the energy density $E/L^p \equiv \varrho$,   regimes of
instability and negative specific heat for $p<7$.
 
 If $p<5$, the microcanonical and canonical ensembles agree for low energy
density,  
$
\varrho\ll \varrho_s
$, 
with $\varrho_s = (\ls)^{-p-1}$ the string scale density. On the other
hand, for high density,  
$
\varrho\gg \varrho_s 
$, 
the specific heat is negative and we enter the long-string phase\foot{It is
unclear whether the long open strings attached to the brane should be
thought of as tracing a random walk in the world-volume, or exploring
the full ten-dimensional ambient space.}, analogous to
the standard behaviour of closed strings in infinite volume. If we agree to
call `Hagedorn regime' the one with super-stringy energy density, we find that
the Hagedorn regime of open strings on the $p<5$ brane is similar to the
one of closed strings and typical of gravitational systems.
 
For $p=5$ the condition for the canonical and microcanonical ensembles to
be equivalent is
$$
{\varrho \over \varrho_s} \ll {\rm log}\left({\varrho\over\varrho_s}\right) +
{\rm log}\left({L \over \ls}\right)
.$$
where $L$ is the length scale of the spatial world-volume. 
This is always satisfied for macroscopic volumes provided we keep the energy
density fixed in the thermodynamical limit. So, very dense systems at finite
volume would be unstable and get negative specific heat, but this does not
occur for sufficiently large world-volume.
 
For $p=6$ the condition of equivalence of ensembles becomes
$$
{\varrho\over \varrho_s} \ll \left({L \over \ls}\right)^p \,,  
$$
which is also easily attainable for sufficently large volume.
For $p\geq 7$ the canonical and microcanonical ensembles are equivalent without
restrictions.
 
Thus, the microcanonical analysis seems to be in agreement with the view that
$p=5$ is the critical dimension above which (at least for large world-volume)
the Hagedorn regime has a `standard' thermodynamics.
  This  critical dimension has been singled out by both considerations
of one-loop holography, and the  Hagedorn cernsorship tests.
This  coincidence  suggests that  the
dualities of \refs\rmalda\ may have a generalization to extreme situations,
such as stringy energy densities.
The main obstacle, in 
implementing any such scenario,  might be the apparent
ten-dimensional behaviour that one finds when the supergravity description
is  pushed to very high temperatures. When the temperature at the
`neck' is of the order of the string scale, the full throat is cut-off
by Hagedorn effects, and there seems to be no effective way of separating
the bulk ten-dimensional dynamics from the `world-volume' dynamics. The
considerations in this section, showing that, at very weak coupling,
$Dp$-branes with $p<5$ have a Hagedorn phase similar to that of closed strings,
could suggest that these two systems are related to each other without
further phase transitions. Roughly speaking, the highly excited long strings
would be random walks in ten-dimensions, even if their end-points are
confined on the $p+1$ -dimensional world-volume.  

Even if we keep $N$ finite, and consider a weak, though finite, coupling
between the open strings on the brane and the closed strings in the
bulk, the thermodynamics of the combined system  may be rather non-trivial. 
 Since   the
Hagedorn temperature can be either limiting or achievable for different
branes,    
depending on $p$, the whole system either may
   reach the phase transition point, or (if $p \geq 5$) the brane will
 take all the energy from the bulk and  stop the whole system
 from reaching the Hagedorn temperature. Other interesting
situations can occur when we have branes  inside branes. We will not  
discuss this subject further here,   and we hope to return to it in a
 future publication.

\newsec{Discussion}

One of the advantages of considering the supergravity limit of the
AdS/CFT correspondence is the access it allows to some non-perturbative
information concerning the systems. This is not yet readily available
in more direct stringy framework. In this note we have attempted to
extract non-perturbative aspects related to the existence of
stringy scales. We have done that by both arranging circumstances
in which the system is challenged to exhibit its non-perturbative nature
at string scales, and also by searching for signs that the correspondence
contains its own limitations. 

By considering a toroidal space-time world-volume for the appropriate
field theory, we have set up a situation where the corresponding
supergravity manifolds contain space and time string scales for low
values of the field-theory temperature. We have then motivated the
way in which we suggest that dynamical effects modify the manifolds
at the string scale.  Manifolds containing string-scale temporal   
loops were capped by a ``stretched horizon", while T-duality considerations
were applied to those containing small spatial loops. Based on this
qualitative picture we  have analyzed the consequences of the 
correspondence. For the ``vacuum" manifolds with stringy temporal loops,
one could have expected Hagedorn effects to take place locally in the
throat part of the manifold. Moreover, that opened the possibility that
Hagedorn behaviour of closed strings could have been computable on
the gauge theory side. We have found however that smooth black-hole
manifolds containing no string scales are favoured upon the ``vacuum"
manifolds, and thus they censor the probing of the Hagedorn regime.

In string theory, one would imagine that the string would be asked to
propagate perturbatively on the ``vacuum" manifold and thus the 
censorship could indicate that in these special backgrounds the
Hagedorn transition is just not there, once the fully fledged non-perturbative
string theory is considered.               
This picture could be reinforced by noting that straining the parameters
in a manner that the Hagedorn stretched horizon does 
 dominate over the standard
smooth horizon, leads at the same time to a spatial manifold with a large
bulk curvature (appropriate to the string scale.) The string spectrum in
such a manifold cannot be estimated by semi-classical methods, and eventually
one would presumably find that the Hagedorn behaviour sets in when the 
boundary temperature itself is again of the string scale.
We have also presented a scaling argument showing that, on the supergravity
side, the string scale enters the physics only in that combination that can
be re-expressed in terms of the coupling of the gauge theory. 

Manifolds with smooth bulk curvature and Hagedorn thresholds can dominate
thermodynamics only for $p>4$, but in those cases we have shown that
holography breaks down, (at least in a naive sense,) and it is difficult
to localize the system in $p+1$ dimensions. A ``conspiracy", taking
place in the CFT/AdS correspondence and its cousins is at work. The
various closed-string/gravity dynamical features always manage to reproduce
non-gravitational, gauge-theory physics, when reinterpreted in terms of
the boundary theory.
The breakdown of holography for $p>4$ is a case where the system had indicated
its own limitations. 

We have searched for further limitations by considering manifolds which are
not fully blown-up, but rather contain a finite throat merging into flat
ten-dimensional Minkowski space. We have found by the similar scaling
arguments mentioned before that, in this case, not all the physics on the
supergravity side could be expressed in terms of the gauge coupling. This
could mean that, while there is still a correspondence in this situation,
it is now a correspondence between such supergravity manifolds and gauge
systems containing Dirac--Born--Infeld corrections. The possibility of such
open-closed string correspondence is made more feasible by a special feature  
of the thermodynamics of weakly coupled open strings on D-branes. It turns out
that for $p<5$ (once again, $p=5$ being a border-line case,) open strings
on Dp-branes have characteristics usually associated with closed strings
propagating in the absence of branes: at Hagedorn temperatures they may exhibit
a phase transition.

The consideration of the thermodynamics of such manifolds leads to uncovering
a rich qualitative structure, some of which could perhaps serve as a basis
for a phase diagram of some gravitational systems. It is encouraging
that the different regions emerging have their counterparts in  
the string/black-hole correspondence principle of \refs\rcorrp. The identification
of such non-perturbative features directly in a string framework remains
a chalenge, and knowing the answer in some cases may be of some help.

\newsec{Aknowledgements}  
We thank  D. Birmingham, 
M.B. Green, D. Kutasov and  J.M. Maldacena  
for useful discussions.  The work of E. R. is partially
supported by the Israel Academy of Sciences and Humanities--Centres
of Excellence Programme, and the American--Israel Bi-National Science
Foundation.

\listrefs
 

\bye